# Dynamic and stochastic systems as a framework for metaphysics and the philosophy of science

Christian List and Marcus Pivato[1]

16 March 2015, with minor changes on 13 August 2015

**Abstract:** Scientists often think of the world (or some part of it) as a dynamical system, a stochastic process, or a generalization of such a system. Prominent examples of systems are (i) the system of planets orbiting the sun or any other classical mechanical system, (ii) a hydrogen atom or any other quantum-mechanical system, and (iii) the earth's atmosphere or any other statistical mechanical system. We introduce a simple and general framework for describing such systems and show how it can be used to examine some familiar philosophical questions, including the following: how can we define nomological possibility, necessity, determinism, and indeterminism; what are symmetries and laws; what regularities must a system display to make scientific inference possible; is there any metaphysical basis for invoking principles of parsimony such as Occam's Razor when we make such inferences; and what is the role of space and time in a system? Our framework is intended to serve as a toolbox for the formal analysis of systems that is applicable in several areas of philosophy.

## 1. Introduction

For both scientific and philosophical purposes, we often find it useful to think of the world (or some part of it that we are studying) as a system evolving over time: a dynamical system, a stochastic process, or a suitable generalization of such a system. In both science and philosophy, many theories represent the world (or the part they are concerned with) in terms of such systems, with various structures and properties. Metaphysical commitments often take the form of claims about the nature of those structures and properties: which of them are real and not just artefacts of our models, which are fundamental as opposed to derivative, and which are necessary as opposed to contingent.

In this paper, we introduce a simple and general framework for describing systems, based on the theory of dynamical systems and stochastic processes, and show how this framework can be used to examine and illuminate some familiar philosophical questions. Here are some examples:

- What does it mean for a system to be deterministic or indeterministic, and which features of the system, if any, determine which others?

- Does the present determine the future? Does it determine the past? What is the smallest set of facts encoding the system's entire history?

- How can we define nomological possibility and necessity for a system?

---

[1] C. List, Departments of Government and Philosophy, London School of Economics, U.K.; M. Pivato, THEMA, Université de Cergy-Pontoise, France. Part of this paper was written when Marcus Pivato was at the Department of Mathematics at Trent University, Canada. Christian List's work was supported by a Leverhulme Major Research Fellowship, and Marcus Pivato's work was supported by an NSERC grant #262620-2008 and also by Labex MME-DII (ANR11-LBX-0023-01). We are very grateful to George Musser and Bryan Roberts for helpful comments.

- What are the laws governing a particular system, and is there a distinction between laws and "brute facts"? How do laws depend on symmetries in a system?

- How much regularity must a system display in order to permit generalizations from local observations to global laws?

- Is there any metaphysical basis for invoking principles of parsimony such as Occam's Razor when we make such generalizations? And how can we formulate Occam's Razor precisely?

- What is the role of space and time in a system? What is the relationship between the geometry of space and time and the system's behaviour?

- Is this spatiotemporal geometry exogenous, or is it somehow determined by the dynamics? In other words, are space and time more fundamental than the system's dynamics, or the other way round?

- How should we individuate systems? Should two structurally indistinguishable systems count as "the same", or might they count as different?

For each of these questions, our framework allows us to identify in simple terms what is at stake. We illustrate the generality of the framework by sketching how it can accommodate, schematically, the systems described by some standard physical theories, such as classical mechanics, electrodynamics, quantum mechanics, and special and general relativity. In principle, our framework can also be used to describe many systems studied in the special sciences, such as biological, social, and economic systems, though we do not have the space to develop these applications here. We make a few remarks about special-science systems at the end of the paper and hope that our framework will serve as a basis for future work in some of those areas.

The paper is structured as follows. We discuss three classes of systems, in increasing order of generality. We call the first *temporally evolving systems* (Section 2), the second *spatially extended systems* (Section 3), and the third *amorphous systems* (Section 4). We offer a conceptual toolbox for describing and analysing each class of systems, covering notions such as states and histories, determinism and indeterminism, nomological possibility and necessity, constraints and probabilistic properties, symmetries and laws, ergodicity and its significance in making scientific inference possible, Occam's Razor, and the role of space and/or time. We first explain all of these notions in the context of the simplest class of systems (in Section 2) and then generalize from there (in Sections 3 and 4).

Although the paper presupposes a willingness to engage with technical material – and a basic familiarity with science will be helpful – our goal is to keep the exposition as self-contained and accessible as possible. One of the paper's intended contributions is a pedagogical one: to present an accessible framework for the analysis of many of the systems studied in the sciences. Readers with a background in mathematics or physics will no doubt recognize much of the conceptual apparatus that we deploy, and parts of this apparatus appear, for instance, in works by van Fraassen (1989), Berkovitz, Frigg, and Kronz (2006, 2011), Werndl (2009a, 2009b), and Butterfield (2012). However, while many of the ideas originate from the theory of dynamical systems and stochastic processes, we adapt and extend them for our purposes and systematize them in a way that is congenial for exploring metaphysical questions. This, in turn, yields a number of insights, for example concerning (i) the role of symmetries in distinguishing between laws and mere "brute facts", (ii) the significance of ergodicity as a justification for scientific inference, (iii) the relationship between Occam's Razor and the symmetries in a system, and (iv) the possibility that the topology and geometry of space and time may be emergent properties of the correlation structure in a system.

**2. Temporally evolving systems**

*2.1 Basic definitions*

We begin with the simplest class of systems whose states evolve over time.[2] *Time* is represented by a set of points $T$ that is linearly ordered; we write < for the "before" relation. The *state* of such a system at each point in time is given by an element of some *state space* $X$. For the moment, we make no assumptions about the internal structure of the states in $X$; they are uninterpreted primitives. A *history* of the system is a path through the state space, represented by a function $h$ from $T$ into $X$. For each time $t$ in $T$, $h(t)$ is the state of the system at time $t$. In a physical system, each state might, for instance, be a completely specified microphysical state in which the system could be at a particular point in time, and histories would be possible trajectories of the system through its state space over time.

We write $\Omega$ to denote the set of all histories deemed *possible*. Histories play the role of possible worlds. Thus the structure of $\Omega$ reflects the notion of possibility we wish to capture. If we are interested in logical possibility, then $\Omega$ is simply the set of all logically possible functions from $T$ into $X$, which we call $\mathcal{H}$. If we are interested in some form of nomological possibility, such as physical possibility, $\Omega$ will often be a proper subset of $\mathcal{H}$. Unless otherwise specified, we adopt the *nomological* interpretation of possibility, since we wish to distinguish between histories that are permitted by the laws governing our system and histories that are not.

Subsets of $\Omega$ are called *events*. We can apply logical operations to events. The *conjunction* of two events $E$ and $E'$ is given by their intersection $E \cap E'$. The *disjunction* of two events $E$ and $E'$ is given by their union $E \cup E'$. The *negation* of an event $E$ is given by its *complement* $\sim E = \Omega \backslash E$. Later we introduce *possibility* and *necessity* operators.

To complete the formal definition of a temporally evolving system, we need to define probabilities on $\Omega$. Formally, we introduce a *conditional probability structure*.[3] This is a family of conditional probability functions $\{Pr_E\}_{E \subseteq \Omega}$, consisting of one $Pr_E$ for each event $E$ in $\Omega$, where $Pr_E$ assigns to any event in $\Omega$ the conditional probability of that event, given $E$.[4] The family must satisfy certain consistency conditions, such as compatibility with Bayesian conditionalization.[5] A *temporally evolving system* is the pair consisting of the set $\Omega$ of possible histories and the conditional probability structure $\{Pr_E\}_{E\subseteq\Omega}$.

---

[2] We build on the formalism in List (2014) and List and Pivato (2015).
[3] Conditional probability structures have previously been considered by several authors, e.g., Popper (1968), Renyi (1955), van Fraassen (1976), as reviewed in Halpern (2010).

For example, in a weather system, $X$ would be the set of all possible weather states and $\Omega$ the set of all possible weather histories. For each particular weather event $E$, say a hot temperature on Wednesday, the function $Pr_E$ then assigns to every weather event $D$, say a thunderstorm on Thursday, the conditional probability of its occurrence, given $E$.

In principle, the probability structure admits two interpretations. Under an *objectivist* interpretation, it is a feature of the system itself and thus represents *objective chance* (see, e.g., Lewis 1986, Schaffer 2007, and List and Pivato 2015). Of course, objective chance could be *degenerate*, i.e., restricted to the extremal values 0 or 1. Degenerate objective chance is a much-discussed feature of deterministic systems; we return to this point later. Under an alternative, *subjectivist* interpretation, the probability structure is not a feature of the system itself, but represents an observer's beliefs about the system, as in subjective Bayesianism (e.g., de Finetti 1972). The most natural way to read this paper is to assume the objectivist interpretation, even though our mathematical analysis does not depend on this.

Familiar examples of temporally evolving systems are (i) the system of planets orbiting the sun or any other classical mechanical system, (ii) a hydrogen atom or any other quantum-mechanical system, (iii) the earth's climate system or any other statistical mechanical system, and (iv) (arguably) the global economy or some other closed macroeconomic system. Generally, any classical *dynamical system*, as standardly defined, is a special case of a temporally evolving system.[6]

For theoretical simplicity, we focus on *closed* systems, which are not subject to any external perturbations. However, one could also represent *open* systems in our framework, by encoding any external perturbations as additional sources of randomness in the system's conditional probability structure ("random forcings").[7]

---

[4] Formally, each $Pr_E$ is defined on a suitable $\sigma$-algebra on $\Omega$; we set the technicalities aside. For any non-empty $E$, $Pr_E$ has all the standard properties of a probability measure. However, for technical reasons, $Pr_\varnothing(D) = 1$ for all $D$.

[5] To be precise, for any subsets $C \subseteq D \subseteq E \subseteq \Omega$, we have $Pr_E(C) = Pr_E(D) \times Pr_D(C)$. Also, $Pr_E(E) = 1$ for all $E \subseteq \Omega$.

[6] A classical *dynamical system* consists of a set $X$ (the *state space*) and a function $\phi$ from $X$ into itself that determines how the state changes over time (the *dynamics*). Let $T=\{0,1,2,3,....\}$. Given any state $x$ in $X$ (the *initial conditions*), the *orbit* of $x$ is the history $h$ defined by $h(0)=x$, $h(1)=\phi(x)$, $h(2)=\phi(\phi(x))$, and so on. Let $\Omega$ be the set of all orbits determined by $(X, \phi)$ in this way. Let $\{Pr'_E\}_{E\subseteq X}$ be any conditional probability structure on $X$. For any events $E$ and $D$ in $\Omega$, we define $Pr_E(D) = Pr'_{E'}(D')$, where $E'$ is the set of all states $x$ in $X$ whose orbits lie in $E$, and $D'$ is the set of all states $x$ in $X$ whose orbits lie in $D$. Then $\{Pr_E\}_{E\subseteq\Omega}$ is a conditional probability structure on $\Omega$. Thus, $\Omega$ and $\{Pr_E\}_{E\subseteq\Omega}$ together form a temporally evolving system. However, not every temporally evolving system arises in this way. In Sections 3 and 4, we extend our framework to even more general classes of systems.

[7] The use of such random forcings does not imply that certain features of the world are genuinely random. Instead, the "randomness" of such forcings is best understood *epistemically* – as a shortcut for an explicit and detailed description of the part of the world which lies outside the model. (This is true whether we adopt an *objectivist* or *subjectivist* interpretation of the probability structure overall.)

*2.2 Determinism and indeterminism*

Conventionally, a system is called *deterministic* if, in that system, the past always determines the future. Formally, for any history $h$ and any point in time $t$, let $h_t$ be the *initial segment* of that history up to $t$. This is the function $h$ restricted to the points in time up to $t$. The system is *deterministic* if, for any history $h$ in $\Omega$ and any time $t$ in $T$, the initial segment $h_t$ admits only one possible continuation in $\Omega$, where a *continuation* of $h_t$ is a history $h'$ such that $h'_t = h_t$. The system is *indeterministic* if, for some $h$ and some $t$, $h_t$ has more than one possible continuation in $\Omega$.[8]

For example, classical mechanical systems, such as the solar system on the Newtonian picture, are deterministic. By contrast, quantum-mechanical systems, such as a decaying uranium atom, are indeterministic (assuming no hidden variables). If the wave function, which encodes the state of the quantum system, collapses at time $t$, the initial segment $h_t$ of the system's history $h$ can admit multiple continuations.

Indeterministic systems, unlike deterministic ones, allow non-degenerate chance even as we move along a given history.[9] Let $E$ be the event that the initial segment $h_t$ has occurred. Then $E$ is formally the set of all continuations of $h_t$. If the system is deterministic, the conditional probability function $Pr_E$ is always *degenerate*, i.e., it assigns probability 0 or 1 to every event $D$. This is because, under determinism, the initial segment $h_t$ has only one continuation, and so the event $E$ that we have defined contains only a single history. Then $Pr_E(D)$ is 1 if that history belongs to $D$ and 0 otherwise. In contrast, if the system is indeterministic, then $Pr_E$ may be *non-degenerate*, assigning probabilities strictly between 0 and 1 to some events $D$. This is because $E$ need not be singleton here, and so $Pr_E$ is less constrained. (For the moment, we set aside phenomena such as "higher-level" indeterminism and chance, as discussed in List and Pivato 2015. We make a few remarks about such phenomena at the end of this paper.)

More generally, our framework allows us to formulate different notions of determinism. For *any* subset $T'$ of $T$, we can ask whether the restriction of a given history to the points in $T'$ uniquely determines the rest of that history. Let $h_{T'}$ denote the restriction of the function $h$ to $T'$. Our question then becomes whether $h_{T'}$ has a unique extension to all of $T$ in $\Omega$, where an *extension* of $h_{T'}$ is a history $h'$ such that $h'_{T'} = h_{T'}$. The set of points in time up to a particular time $t$ is just one special case of what the set $T'$ might be.

In this way, we might ask, for instance, whether the complete history of a system, both past and future, is determined by its present state alone. Similarly, we might ask whether, given the states of the system at two points in time, there is a unique history connecting them. So, one can, in principle, consider not only the widely discussed idea of "past-to-future" determinism, but also various other forms of "local-to-global" determinism.

---

[8] On these definitions, see also List (2014) and List and Pivato (2015).
[9] For discussion, see, e.g., Schaffer (2007) and List and Pivato (2015).

*2.3 Nomological possibility and necessity*

We can explicitly define the notions of nomological necessity and possibility in our framework.[10] Intuitively, an event $E$ is nomologically possible in history $h$ at time $t$ if the initial segment of that history up to $t$ admits *at least one* continuation in $\Omega$ that lies in $E$; and $E$ is nomologically necessary in $h$ at time $t$ if *every* continuation of the history's initial segment up to $t$ lies in $E$.

More formally, we say that one history, $h'$, is *accessible* from another, $h$, at time $t$ if the initial segments of $h$ and $h'$ up to time $t$ coincide, i.e., $h_t = h_t'$. We then write $h'R_t h$. The binary relation $R_t$ on possible histories is in fact an equivalence relation (reflexive, symmetric, and transitive). Now, an event $E \subseteq \Omega$ is *nomologically possible* in history $h$ at time $t$ if *some* history $h'$ in $\Omega$ that is accessible from $h$ at $t$ is contained in $E$. Similarly, an event $E \subseteq \Omega$ is *nomologically necessary* in history $h$ at time $t$ if *every* history $h'$ in $\Omega$ that is accessible from $h$ at $t$ is contained in $E$.

In this way, we can define two modal operators, $\blacklozenge_t$ and $\blacksquare_t$, to express possibility and necessity at time $t$. We define each of them as a mapping from events to events. For any event $E \subseteq \Omega$,

$\blacklozenge_t E = \{h \in \Omega : \text{for some } h' \in \Omega \text{ with } h'R_t h, \text{ we have } h' \in E\}$,

$\blacksquare_t E = \{h \in \Omega : \text{for all } h' \in \Omega \text{ with } h'R_t h, \text{ we have } h' \in E\}$.

So, $\blacklozenge_t E$ is the set of all histories in which $E$ is possible at time $t$, and $\blacksquare_t E$ is the set of all histories in which $E$ is necessary at time $t$. Accordingly, we say that "$\blacklozenge_t E$" holds in history $h$ if $h$ is an element of $\blacklozenge_t E$, and "$\blacksquare_t E$" holds in $h$ if $h$ is an element of $\blacksquare_t E$. As one would expect, the two modal operators are *duals* of each other: for any event $E \subseteq \Omega$, we have $\blacksquare_t E = \sim\blacklozenge_t \sim E$ and $\blacklozenge_t E = \sim\blacksquare_t \sim E$.

Two remarks are due. First, although we have here defined *nomological* possibility and necessity, we can analogously define *logical* possibility and necessity. To do this, we must simply replace every occurrence of the set $\Omega$ of nomologically possible histories in our definitions with the set $\mathcal{H}$ of logically possible histories. Second, by defining the operators $\blacklozenge_t$ and $\blacksquare_t$ as functions from events to events, we have adopted a *semantic* definition of these modal notions. However, we could also define them *syntactically*, by introducing an explicit modal logic. For each point in time $t$, the logic corresponding to the operators $\blacklozenge_t$ and $\blacksquare_t$ would then be an instance of a standard S5 modal logic (on S5, see, e.g., Priest 2001).

Our analysis shows how nomological possibility and necessity depend on the dynamics of the system. In particular, as time progresses, the notion of possibility becomes more demanding: fewer events remain possible at each time. And the notion of necessity becomes less demanding: more events become necessary at each time, for instance due to having been "settled" in the past. Formally, for any $t$ and $t'$ in $T$ with $t < t'$ and any event $E \subseteq \Omega$, if $\blacklozenge_{t'} E$ then $\blacklozenge_t E$,

---

[10] We here employ a construction from List (2014).

if ■$_t E$ then ■$_{t'} E$.

Furthermore, in a deterministic system, for every event $E$ and any time $t$, we have ◆$_t E$ = ■$_t E$. In other words, an event is possible in any history $h$ at time $t$ if and only if it is necessary in $h$ at $t$. In an indeterministic system, by contrast, necessity and possibility come apart.

Just as we previously discussed different notions of determinism – not just "past to future" but also "local to global" – so we can generalize the notions of possibility and necessity in a similar way. Let us say that one history, $h'$, is *accessible* from another, $h$, relative to a set $T'$ of time points, if the restrictions of $h$ and $h'$ to $T'$ coincide, i.e., $h'_{T'} = h_{T'}$. We then write $h'R_T h$. Accessibility at time $t$ is the special case where $T'$ is the set of points in time up to time $t$. We can define nomological possibility and necessity relative to $T'$ as follows. For any event $E \subseteq \Omega$,

◆$_{T'} E$ = {$h \in \Omega$ : for some $h' \in \Omega$ with $h'R_T h$, we have $h' \in E$},

■$_{T'} E$ = {$h \in \Omega$ : for all $h' \in \Omega$ with $h'R_T h$, we have $h' \in E$}.

Although these modal notions are much less familiar than the standard ones (possibility and necessity at time $t$), they are useful for some purposes. In particular, they allow us to express the fact that the states of a system during a particular period of time, $T' \subseteq T$, render some events $E$ possible or necessary.

Finally, our definitions of possibility and necessity relative to some general subset $T'$ of $T$ also allow us to define completely "atemporal" notions of possibility and necessity. If we take $T'$ to be the empty set, then the accessibility relation $R_{T'}$ becomes the universal relation, under which every history is related to every other. An event $E$ is possible in this atemporal sense (i.e., ◆$_\varnothing E$) if and only if $E$ is a non-empty subset of $\Omega$, and it is necessary in this atemporal sense (i.e., ■$_\varnothing E$) if $E$ coincides with all of $\Omega$. These notions might be viewed as possibility and necessity from the perspective of some observer who has no temporal or historical location within the system and looks at it from the outside.

*2.4 Constraints and correlations*

Ultimately, all modal constraints governing a temporally evolving system are encoded by the set $\Omega$, and all correlations, or probabilistic properties, are encoded by the conditional probability structure {$Pr_E$}$_{E \subseteq \Omega}$. This raises the question: which of these constraints and correlations qualify as "laws", rather than as mere "brute facts" about the system? Further, is the distinction between laws and brute facts even meaningful?

Intuitively, we take laws to be constraints or correlations that, in some sense, hold universally, as distinct from constraints or correlations that are specific to particular circumstances. How can we make sense of this idea in relation to a temporally evolving system? After all, the constraints encoded by $\Omega$ and the correlations encoded by {$Pr_E$}$_{E \subseteq \Omega}$ seem themselves to be "timeless". From a bird's-eye perspective, they govern the behaviour of the system in its entirety. Can we still draw a distinction between these "overall" features of the system and the laws more specifically?

In what follows, we introduce two general notions – constraints on histories and probabilistic properties – and in each case provide a criterion for identifying which of them qualify as laws. Informally, a *constraint* on histories is a property, *C*, that a history may or may not have. Formally, *C* can be associated with some subset, denoted [*C*], of the set $\mathcal{H}$ of all logically possible histories. A history satisfies the constraint *C* if it belongs to [*C*]. We call [*C*] the *extension* of *C*. The system as a whole satisfies *C* if *all* histories in Ω satisfy *C*, i.e., if Ω is a subset of [*C*]. For example, each of Newton's three laws of motion can be viewed as a constraint that all histories of a classical mechanical system satisfy. Note that, since the set Ω of all nomologically possible histories is itself a subset of the set $\mathcal{H}$ of all logically possible histories, we can think of Ω as the extension of the "total nomological constraint" governing our system.

Let us turn to probabilistic properties. A *probabilistic property*, *P*, is a property that a conditional probability structure may or may not have. Formally, it is associated with a subset, denoted [*P*], of the set Π of all logically possible conditional probability structures on Ω. A conditional probability structure $\{Pr_E\}_{E \subseteq \Omega}$ satisfies *P* if it belongs to [*P*]. We call [*P*] the *extension* of *P*. For example, we can think of the second law of thermodynamics as a property of the conditional probability structure of a statistical mechanical system.

Our goal is to formulate conditions that allow us to distinguish between constraints or probabilistic properties that "hold universally" and ones that do not. The former can be viewed as "laws", the latter merely as "brute facts". We capture this distinction through the notion of *symmetries*. Informally, a *symmetry* is a special kind of transformation that acts on either the state space *X* or the set of time points *T* or both. We can then define *laws* as those constraints or probabilistic properties that are invariant under relevant symmetries, i.e., whose extensions are preserved by them. We now make these notions formally precise.

*2.5 Symmetries*

We first consider symmetries acting on the state space; we then turn to symmetries acting on time; and we finally consider more general symmetries. (To anticipate: in the most general case, a symmetry is a function that acts on the set of logically possible histories in such a way as to preserve certain significant features.)

To introduce *state symmetries*, let us begin with some preliminary definitions. Let ϕ be any function from *X* into itself, i.e., a transformation on the state space. We can use this transformation to define a function from histories to other histories. Specifically, for any history *h*, we define the transformed history

$$\phi(h) = h', \text{ where, for all } t \text{ in } T, h'(t) = \phi[h(t)].$$

For example, suppose *T*={1,2,3,…} and *X*={*a,b,c,d,...,z*}. Here, any history *h* can be represented as a sequence of elements in *X*. Suppose ϕ is the function that shifts every letter in the alphabet one place to the right, i.e., *a* to *b*, *b* to *c*, and so on, and *z* back to *a*. Then applying ϕ to the history *h* = (*b,a,c,f,z,…*) yields the history *h'*= (*c,b,d,g,a,…*). Note that, even when *h* is in Ω, its image *h'* might not be an element of Ω. It is simply some logically possible function from *T* into *X*. Thus ϕ induces a function from the set $\mathcal{H}$ of logically possible histories to itself. For convenience, we use the letter ϕ both to denote the original function on $X$ and to denote the induced function on $\mathcal{H}$.

Given any event $E$ in Ω (or more generally any subset $E$ of $\mathcal{H}$), we define the *inverse image* of $E$ under ϕ to be the set of all histories $h$ in $\mathcal{H}$ such that ϕ($h$) lies in $E$.[11] For example, if $E$ is the set of all histories whose state at time 3 is $c$, then the inverse image of $E$ under ϕ is the set of all histories whose state at time 3 is $b$. Note that the inverse image of an event $E$ could be empty, namely if none of the histories in $E$ can be "reached" as transformations of other histories.

The function ϕ is a *symmetry* of our system if

- ϕ($h$) is in Ω, for all $h$ in Ω; and
- for any events $E$ and $D$ in Ω, if $E'$ and $D'$ are the inverse images of $E$ and $D$ under ϕ, then $Pr_{E'}[D'] = Pr_E[D]$.[12]

Intuitively, a symmetry is a transformation that preserves the modal and probabilistic structure of a temporally evolving system. In our example, where $X=\{a,b,c,d,...,z\}$ and ϕ is the letter-shifting function, the first part of this definition implies that if ($b,a,c,f,z,...$) is a nomologically possible history of the system, then so is ($c,b,d,g,a,...$). To illustrate the second part, let $E$ be the set of all histories in Ω whose state at time 3 is $c$, and let $D$ be the set of all histories in Ω whose state at time 5 is $a$ (so that $E'$ is a suitable set of histories whose state at time 3 is $b$, and $D'$ is a suitable set of histories whose state at time 5 is $z$).[13] The conditional probability that the state of a history at time 5 is $a$, given that at time 3 it is $c$, must then equal the conditional probability that the state at time 5 is $z$, given that at time 3 it is $b$.

Obviously, not all state transformations are symmetries. Whether there are any non-trivial state symmetries and, if so, what they look like, depends very much on the temporally evolving system in question, i.e., it depends on the set Ω and the conditional probability structure $\{Pr_E\}_{E \subseteq \Omega}$. In classical mechanical systems, state symmetries include spatial translations, which shift everything in a certain direction by a certain distance, rotations and reflections, and permutations of particles with equal mass. Those transformations preserve the modal and probabilistic structure of the systems in question.

Similarly, we can define symmetries acting on time. Again, we begin with some preliminary definitions. Let ψ be any function on $T$, i.e., a transformation on time. For any history $h$, we define the transformed history

---

[11] Formally, we write $\phi^{-1}(E) = \{h$ in $\mathcal{H}$: ϕ($h$) is in $E\}$. Note that the use of this notation for inverse images of sets does not imply that the function ϕ is invertible.

[12] Strictly speaking, the conditional probability structure $\{Pr_E\}_{E \subseteq \Omega}$ is only defined for subsets of Ω. However, we can extend it to the rest of $\mathcal{H}$ in a straightforward way: for any subsets $D$, $E$ of $\mathcal{H}$, we define $Pr_E(D) = Pr_{E \cap \Omega}(D \cap \Omega)$.

[13] Formally, $E'$ consists of all the histories $h$ in $\mathcal{H}$ such that ϕ($h$) is in Ω and $h(3) = b$. Similarly, $D'$ consists of all the histories $h$ in $\mathcal{H}$ such that ϕ($h$) is in Ω and $h(5) = z$.

$\psi(h) = h'$, where, for all $t$ in $T$, $h'(t) = h[\psi(t)]$.[14]

For example, suppose $T = \{1,2,3,...\}$. Then any history $h$ can be represented as a sequence $(x_1, x_2, x_3, ...)$ of elements in $X$. If $\psi(t) = t + 5$ for all $t$ in $T$, then $\psi(h) = (x_6, x_7, x_8, ...)$. As in the case of state symmetries, $\psi$ induces a function from the set $\mathcal{H}$ to itself. Again, the *inverse image* of any set $E$ of histories under $\psi$ is the set of all histories $h$ in $\mathcal{H}$ such that $\psi(h)$ lies in $E$.[15]

In analogy to the earlier definition, the function $\psi$ is a *symmetry* if

- $\psi(h)$ is in $\Omega$, for all $h$ in $\Omega$; and
- for any events $E$ and $D$ in $\Omega$, if $E'$ and $D'$ are the inverse images of $E$ and $D$ under $\psi$, then $Pr_{E'}[D'] = Pr_E[D]$.

In our example, where $T = \{1,2,3,...\}$ and $\psi(t) = t + 5$, the first part of this definition says that if $h = (x_1, x_2, x_3,...)$ is a nomologically possible history of the system, then so is $h' = (x_6, x_7, x_8,...)$. To illustrate the second part, suppose that $E$ is the set of all histories in $\Omega$ whose state at time 3 is $c$, while $D$ is the set of all histories in $\Omega$ whose state at time 4 is $a$ (so that $E'$ is a suitable set of histories whose state at time 8 is $c$, while $D'$ is a suitable set of histories whose state at time 9 is $a$). The conditional probability that the state in a history at time 9 is $a$, given that at time 8 it was $c$, must then equal the conditional probability that the state at time 4 is $a$, given that at time 3 it was $c$.[16]

Just as it is not the case that all state transformations are symmetries, so it is not the case that all time transformations are symmetries. In most systems arising in classical physics, time symmetries include *time translations*, such as $\psi(t) = 5+t$, but exclude *non-linear transformations*, such as $\psi(t) = t^2$. In systems where the state does not encode explicitly "kinetic" properties (such as momentum), *simple time reversals*, such as $\psi(t) = -t$, can also be time symmetries. For example, the partial differential equations describing wave propagation in an ideal medium are invariant under simple time reversals. However, many other systems, such as thermodynamic systems and diffusion processes, do not admit time-reversal symmetries.

---

[14] Typically, we require $\psi$ to be *order-preserving*, i.e., for all $t$ and $t'$ in $T$, if $t < t'$, then $\psi(t) < \psi(t')$. For example, if $T = \{1,2,3,...\}$ with the standard ordering, the functions $\psi(t) = t + 5$ and $\psi(t) = 5t$ are order-preserving. But we do not build this requirement into our definition of a time symmetry. Note that some time symmetries, such as time reversals in classical physical systems, are not order-preserving.

[15] Formally, we write $\psi^{-1}(E) = \{h$ in $\mathcal{H}: \psi(h)$ is in $E\}$. Again, this set could be empty. Suppose, for example, that $T = \{1,2,3,...\}$ and $\psi(t) = t+5$. If $E$ is the set of all histories whose state at time 3 is $c$ (where states, as before, are letters of the alphabet), then the inverse image of $E$ under $\psi$ is the set of all histories whose state at time 8 is $c$. This is because the third state of any history $h$ in $E$ must be the eighth state of some history $h'$ in its inverse image $E'$ under $\psi$.

[16] Note that classical dynamical systems have a particularly rich set of time symmetries. Let $(X, \phi)$ be a dynamical system, as defined in footnote 6. Suppose the function $\phi$ (which maps from $X$ into itself) is *surjective*, i.e., for all $x$ in $X$, there is some $y$ in $X$ such that $\phi(y)=x$. Then the set $\Omega$ of orbits is invariant under all time-shifts. Let $\{Pr'_E\}_{E \subseteq X}$ be a conditional probability structure on $X$, and let $\{Pr_E\}_{E \subseteq \Omega}$ be the conditional probability structure it induces on $\Omega$. Suppose that $\{Pr'_E\}_{E \subseteq X}$ is $\phi$-*invariant*, i.e., for any subsets $E$ and $D$ of $X$, if $E' = \phi^{-1}(E)$ and $D' = \phi^{-1}(D)$, then $Pr'_E(D') = Pr'_E(D)$. Then *every* time shift is a temporal symmetry of the resulting temporally evolving system. The study of dynamical systems equipped with invariant probability measures is the purview of *ergodic theory*.

More general symmetries include composite functions resulting from the combination of transformations of *X* and transformations of *T*. These are best viewed as functions acting on the set Ω of histories directly, with the properties introduced above. A familiar example in classical mechanical systems is a *time reversal*, which involves *both* a negation of the time index *and* a negation of all momentum vectors in the system.[17] A more complex example is a *Galilean transformation*, which adds a constant vector to the momentum vectors of all particles and also a time-varying sequence of spatial shifts to the particle positions, thereby converting the system to a different inertial reference frame. For simplicity, we omit the technical details.

When a transformation of the state space, time, or both is a symmetry of a given system, this encodes the fact that certain properties that hold locally (i.e., at some points in time, at some states, and in some histories) are more general: they are preserved under the transformation in question.

In what follows, we write Γ to denote the set of all symmetries of our temporally evolving system. This set has the algebraic structure of a *monoid*. Formally, a set of transformations of $\mathcal{H}$ is a *monoid* if (i) it contains the *identity transformation* (which maps every history to itself) and (ii) it is *closed under composition* (i.e., for any two transformations in the set, the transformation obtained by applying first one of the two transformations and then the other is also in the set). An example of a monoid of transformations is the set of all rotations of a classical mechanical system around a fixed axis: the identity transformation obviously belongs to this set, being a rotation by an angle of zero, and the composition of any two rotations is still a rotation.

To see that the set Γ of *all* symmetries of a temporally evolving system forms a monoid, note that (i) the identity transformation, which maps every history to itself, is trivially a symmetry, and (ii) if two transformations each qualify as symmetries, by preserving the modal and probabilistic structure of the system, then so does their composition.

*2.6 Laws*

As anticipated, the key difference between laws and features of our system that are just "brute facts" is that the former, but not the latter, are *invariant* under the symmetries of the system. The close relationship between symmetries and laws has of course been recognized by many physicists and philosophers of science (e.g., Wigner 1967; van Fraassen 1989, Part III; Mainzer 1996, Section 5.3; Brading and Castellani 2003, 2013; and Baker 2010). We here offer a formalization of this idea in our framework.

We first consider constraints on histories, which are candidates for *modal laws*, i.e., laws governing what is or is not nomologically possible. We then turn to probabilistic properties, which are candidates for *probabilistic laws*, i.e., laws governing what correlates with what.

---

[17] As Roberts (2013) has argued, in general, a time reversal must not only map the time coordinate *t* to –*t*, but also apply an appropriate transformation to the system's state at each point in time (in special cases, this could be the identity transformation, as in the case of a simple time reversal). Generally, we can think of the state of the system at each point in time as encoding not only some "static" properties (such as each particle's position), but also some "kinetic" properties (such as each particle's momentum). While static properties are preserved under time reversals, kinetic properties are not generally preserved. Similarly, in quantum mechanics, time reversals involve not only a reversal of the time coordinate, but also taking the conjugate of the wave function's values.

Let *C* be a constraint, with extension [*C*], and let γ be a symmetry of the system, i.e., a transformation on $\mathcal{H}$ that belongs to the set Γ. We say that *C* is *invariant* under γ if the set [*C*] is equal to its inverse image under γ. A constraint *C* that is satisfied by the system is a *law* if it is invariant under *all* symmetries in Γ.

For example, suppose *T* = {1,2,3,...} and suppose that, for any positive integer *r*, the set Γ contains the time symmetry $\psi_r$ defined by $\psi_r(t) = t+r$ for all *t* in *T*. Suppose the system satisfies the constraint *C* which says: "if the state of the system at time 5 is *x*, then at time 6 it is *y*". The inverse image of [*C*] under $\psi_2$ corresponds to the constraint *C'* which says: "if the state of the system at time 7 is *x*, then at time 8 it is *y*". Clearly, [*C'*] is not the same as [*C*]. Thus, *C* is *not* invariant under $\psi_2$, and so *C* is not a law of the system. It is simply a constraint that the system happens to satisfy: a "brute fact".

However, suppose the system satisfies a different constraint *C* which says: "for any *t* in *T*, if the state of the system at time *t* is *x*, then at time *t* + 1 it is *y*". It is easy to see that [*C*] *is* invariant under $\psi_r$ for all positive integers *r*. If Γ consists only of the time symmetries {$\psi_r$ : *r* = 1,2,3,....}, then *C* is invariant under *all* elements of Γ. Thus, *C* is a law of the system.

For another example, consider the kinds of temporally evolving systems that arise in classical mechanics. These satisfy the law of *conservation of energy*, which says that the total energy (kinetic plus potential) of the system remains constant over time. This can be formulated as a constraint *C* of the form: "for any times *t* and *t'* in *T*, the total energy of the state at time *t'* equals the total energy of the state at time *t*". Clearly, this constraint is invariant under the time symmetries {$\psi_r$} introduced above. As already mentioned, classical mechanical systems also have certain *state* symmetries, such as spatial translations, rotations, reflections, and the permutation of (equal-mass) particles. The total energy of a state is unchanged by such symmetries, so the constraint *C* will also be invariant under spatial translations and (equal-mass) particle permutations. Indeed, total energy is unchanged by every symmetry of the system, and for this reason, the constraint *C* is a law.[18]

We can also use invariance under symmetries to distinguish between probabilistic laws and "brute facts" about a system's probabilities. Let $\{Pr'_E\}_{E \subseteq \Omega}$ be any conditional probability structure, and let γ be a symmetry of the system. We define $\gamma(\{Pr'_E\}_{E \subseteq \Omega})$ to be the conditional probability structure $\{Pr^*_E\}_{E \subseteq \Omega}$ such that, for any events *E* and *D*, we have $Pr^*_E(D) = Pr'_{E'}(D')$, where *E'* is the inverse image of *E* under γ, and *D'* is the inverse image of *D* under γ. Let *P* be a probabilistic property, and let [*P*] be its extension (thus, [*P*] is a subset of the set Π of all possible conditional probability structures). We say that *P* is *invariant* under γ if [*P*] is equal to its inverse image under γ. A probabilistic property *P* that is satisfied by the conditional probability structure $\{Pr_E\}_{E \subseteq \Omega}$ is a *probabilistic law* if it is invariant under *all* symmetries in Γ.

---

[18] In a similar fashion one can formulate the laws of *conservation of momentum*, *conservation of angular momentum*, and so on.

For example, suppose $T=\{1,2,3,...\}$, and, for any positive integer $r$, define the time symmetry $\psi_r$ as before. Let $Y$ and $Z$ be two subsets of the state space $X$, and suppose the system satisfies the probabilistic property $P$ which says: "conditional on the state being in $Y$ at time 5, there is a 50% probability that the state will be in $Z$ at time 8". The inverse image of $[P]$ under $\psi_2$ corresponds to the probabilistic property $P'$ which says: "conditional on the state being in $Y$ at time 7, there is a 50% probability that the state will be in $Z$ at time 10". Clearly, $[P']$ is not the same as $[P]$. Thus, $[P]$ is *not* invariant under $\psi_2$, and so $P$ is not a probabilistic law of the system.

However, suppose the system satisfies the probabilistic property $P$ which says: "for any time $t$ in $T$, conditional on the state being in $Y$ at time $t$, there is a 50% probability that the state will be in $Z$ at time $t+3$". Then it is easy to see that $[P]$ *is* invariant under $\psi_r$ for all positive integers $r$. If $\Gamma$ consists only of the time symmetries $\{\psi_r : r = 1,2,3,....\}$, then $P$ is invariant under *all* elements of $\Gamma$, and so $P$ is a probabilistic law.

In sum, modal and probabilistic laws, as opposed to "brute facts", are, respectively, those constraints on histories and those probabilistic properties that are preserved by all symmetries of the system. In Appendix A, we extend this account to *factor systems*, which are obtained by abstracting away from certain details of the original system. In Appendix B, we extend it to *partial* and *local* symmetries, which are often found in systems with special initial conditions and/or boundary conditions.

*2.7 From local observations to global laws*

When we scientifically investigate a system, we cannot normally observe all possible histories in $\Omega$, or directly access the conditional probability structure $\{Pr_E\}_{E \subseteq \Omega}$. Instead, we can only observe specific events. Conducting many "runs" of the same experiment is an attempt to observe as many histories of a system as possible, but even the best experimental design rarely allows us to observe *all* histories or to read off the *full* conditional probability structure. Furthermore, this strategy works only for smaller systems that we can isolate in laboratory conditions. When the system is the economy, the global ecosystem, or the universe in its entirety, we are stuck in a single history. We cannot step outside that history and look at alternative histories. Nonetheless, we would like to infer something about the laws of the system in general, and especially about the true probability distribution over histories.

Can we discern the system's laws and true probabilities from observations of specific events? And what kinds of regularities must the system display in order to make this possible? In other words, are there certain "metaphysical prerequisites" that must be in place for scientific inference to work?

To answer these questions, we first consider a very simple example. Here $T = \{1,2,3,...\}$, and the system's state at any time is the outcome of an independent coin toss. So the state space is $X = \{Heads, Tails\}$, and each possible history in $\Omega$ is one possible *Heads/Tails* sequence.

Suppose the true conditional probability structure on $\Omega$ is induced by the single parameter $p$, the probability of *Heads*. In this example, the Law of Large Numbers guarantees that, with probability 1, the limiting frequency of *Heads* in a given history (as time goes to infinity) will match $p$. This means that the subset of $\Omega$ consisting of "well-behaved" histories has probability 1, where a history is well-behaved if (i) there exists a limiting frequency of *Heads* for it (i.e., the proportion of *Heads* converges to a well-defined limit as time goes to infinity) and (ii) that limiting frequency is *p*. For this reason, we will almost certainly (with probability 1) arrive at the true conditional probability structure on $\Omega$ on the basis of observing just a single history and counting the number of *Heads* and *Tails* in it.

Does this result generalize? The short answer is "yes", provided the system's symmetries are of the right kind. Without suitable symmetries, generalizing from local observations to global laws is not possible. In a slogan, for scientific inference to work, there must be sufficient regularities in the system. In our toy system of the coin tosses, there are. Wigner (1967) recognized this point, taking symmetries to be "a prerequisite for the very possibility of discovering the laws of nature" (as Brading and Castellani 2013 put it; see also French 2014).

Generally, symmetries allow us to infer general laws from specific observations. For example, let $T = \{1,2,3,...\}$, and let *Y* and *Z* be two subsets of the state space *X*. Suppose we have made the observation *O*: "whenever the state is in the set *Y* at time 5, there is a 50% probability that it will be in *Z* at time 6". Suppose we know, or are justified in hypothesizing, that the system has the set of time symmetries $\{\psi_r : r = 1,2,3,....\}$, with $\psi_r(t) = t + r$, as defined as in the previous section. Then, from observation *O*, we can deduce the following general law: "for any *t* in *T*, if the state of the system is in the set *Y* at time *t*, there is a 50% probability that it will be in *Z* at time $t + 1$".

However, this example still has a problem. It only shows that *if* we could make observation *O*, *then* our generalization would be warranted, provided the system has the relevant symmetries. But the "if" is a big "if". Recall what observation *O* says: "whenever the system's state is in the set *Y* at time 5, there is a 50% probability that it will be in the set *Z* at time 6". Clearly, this statement is only empirically well supported – and thus a real observation rather than a mere hypothesis – *if* we can make many observations of possible histories at times 5 and 6. We can do this if the system is an experimental apparatus in a lab or a virtual system in a computer, which we are manipulating and observing "from the outside", and on which we can perform many "runs" of an experiment. But, as noted above, if we are participants in the system, as in the case of the economy, an ecosystem, or the universe at large, we only get to experience times 5 and 6 once, and we only get to experience one possible history. How, then, can we ever assemble a body of evidence that allows us to make statements such as *O*?

The solution to this problem lies in the property of *ergodicity*. This is a property that a system may or may not have and that, if present, serves as the desired metaphysical prerequisite for scientific inference. To explain this property, let us give an example. Suppose $T = \{1,2,3,...\}$, and the system has all the time symmetries in the set $\Psi = \{\psi_r : r = 1,2,3,....\}$ (and perhaps other symmetries as well, though we set these aside for now). Heuristically, the symmetries in $\Psi$ can be interpreted as describing the evolution of the system over time.[19] Suppose each time-step corresponds to a day. Then the history $h = (a,b,c,d,e,....)$ describes a situation where today's state is $a$, tomorrow's is $b$, the next day's is $c$, and so on. The transformed history $\psi_1(h) = (b,c,d,e,f,....)$ describes a situation where today's state is $b$, tomorrow's is $c$, the following day's is $d$, and so on. Thus, $\psi_1(h)$ describes the same "world" as $h$, but *as seen from the perspective of tomorrow*. Likewise, $\psi_2(h) = (c,d,e,f,g,....)$ describes the same "world" as $h$, but *as seen from the perspective of the day after tomorrow*, and so on.[20]

Given the set $\Psi$ of symmetries, an event $E$ (a subset of $\Omega$) is $\Psi$-*invariant* if the inverse image of $E$ under $\psi$ is $E$ itself, for all $\psi$ in $\Psi$. This implies that if a history $h$ is in $E$, then $\psi(h)$ will also be in $E$, for all $\psi$. In effect, if the world is in the set $E$ today, it will remain in $E$ tomorrow, and the day after tomorrow, and so on. Thus, $E$ is a "persistent" event: an event one cannot escape from by moving forward in time. In a coin-tossing system, where $\Psi$ is still the set of time translations, examples of $\Psi$-invariant events are "all *Heads*", where $E$ contains only the history (*Heads*, *Heads*, *Heads*, …), and "all *Tails*", where $E$ contains only the history (*Tails*, *Tails*, *Tails*, …).

The system is *ergodic* (with respect to $\Psi$) if, for any $\Psi$-invariant event $E$, the unconditional probability of $E$, i.e., $Pr_\Omega(E)$, is either 0 or 1. In other words, the only persistent events are those which occur in *almost no* history (i.e., $Pr_\Omega(E) = 0$) and those which occur in *almost every* history (i.e., $Pr_\Omega(E) = 1$).[21] Our coin-tossing system is ergodic, as exemplified by the fact that the $\Psi$-invariant events "all *Heads*" and "all *Tails*" occur with probability 0.

In an ergodic system, it is possible to estimate the probability of any event "empirically", by simply counting the frequency with which that event occurs.[22] Frequencies are thus evidence for probabilities. The formal statement of this is the following important result from the theory of dynamical systems and stochastic processes.

> **Ergodic Theorem:** Suppose the system is ergodic. Let $E$ be any event and let $h$ be any history. For all times $t$ in $T$, let $N_t$ be the number of elements $r$ in the set $\{1, 2, ..., t\}$ such that $\psi_r(h)$ is in $E$. Then, with probability 1, the ratio $N_t/t$ will converge to $Pr_\Omega(E)$ as $t$ increases towards infinity.[23]

---

[19] Mathematically, the pair $(\Omega, \Psi)$ can be interpreted as a classical *dynamical system*, as defined in footnote 6, with $\Omega$ playing the role of a state space (from an outside observer's perspective) and the transformations in $\Psi$ playing the role of state transformation rules.

[20] Note that, under this heuristic interpretation, the world "forgets" its past history: from the perspective of tomorrow, it is as if today never happened. This is just an artefact of the formal mathematical model we are using in this example and has no deeper significance. If we used the set **Z** of all integer numbers instead of the natural numbers to model time, it would obviate this issue.

[21] Note that, if $\Omega$ is infinite, there is a subtle distinction between *almost no history* (i.e., $Pr_\Omega(E) = 0$) and *no history* (i.e., $E$ is the empty set). Likewise, there is a subtle distinction between *almost every history* (i.e., $Pr_\Omega(E) = 1$) and *every history* (i.e., $E = \Omega$).

[22] This insight is the basis for Reichenbach's (1949) "straight rule", which is to take observed frequencies as the best estimates of "true" probabilities. See, e.g., Eberhardt and Glymour (2009).

[23] For simplicity, we have stated this result somewhat informally. For formal statements, see Berkovitz, Frigg, and Kronz (2006). Furthermore, we have here defined ergodicity with respect to a set of time

Intuitively, $N_t$ is the number of times the event $E$ has "occurred" in history $h$ from time 1 up to time $t$. The ratio $N_t/t$ is therefore the *frequency* of occurrence of event $E$ (up to time $t$) in history $h$. This frequency might be measured, for example, by performing a sequence of experiments or observations at times 1, 2, ..., $t$. The Ergodic Theorem says that, almost certainly (i.e., with probability 1), the empirical frequency will converge to the *true* probability of $E$, $Pr_\Omega(E)$, as the number of observations becomes large. The estimation of the true conditional probability structure from the frequencies of *Heads* and *Tails* in our illustrative coin-tossing system is possible precisely because the system is ergodic.

To understand the significance of this result, let $Y$ and $Z$ be two subsets of $X$, and suppose $E$ is the event "$h(1)$ is in $Y$", while $D$ is the event "$h(2)$ is in $Z$". Then the intersection $E \cap D$ is the event "$h(1)$ is in $Y$, and $h(2)$ is in $Z$". The Ergodic Theorem says that, by performing a sequence of observations over time, we can empirically estimate $Pr_\Omega(E)$ and $Pr_\Omega(E \cap D)$ with arbitrarily high precision. Thus, we can compute the ratio $Pr_\Omega(E \cap D) / Pr_\Omega(E)$. But this ratio is simply the *conditional probability* $Pr_E(D)$. And so, we are able to estimate the conditional probability that the state at time 2 will be in $Z$, given that at time 1 it was in $Y$. This illustrates that, by allowing us to estimate unconditional probabilities empirically, the Ergodic Theorem also allows us to estimate conditional probabilities, and in this way to learn the properties of the conditional probability structure $\{Pr_E\}_{E \subseteq \Omega}$.

We may thus conclude that ergodicity is what allows us to generalize from local observations to global laws. In effect, when we engage in scientific inference about some system, or even about the world at large, we rely on the hypothesis that this system, or the world, is ergodic.[24] If our system, or the world, were what Cartwright (1999) calls "dappled", then presumably we would not be able to presuppose ergodicity, and hence our ability to make scientific generalizations would be compromised.

*2.8 Occam's Razor*

We have seen that a temporally evolving system must possess a sufficiently rich set of symmetries to allow us to infer general laws from a finite set of empirical observations. In the previous section, we took for granted that we already knew, or at least were justified in hypothesizing, that the system had these symmetries. But what justified this hypothesis?

---

symmetries. More generally, the statements in this section hold for any collection of symmetries that forms an *amenable semi-group*. See Krengel (1985, Section 6.4).

[24] One complication is that not all systems are ergodic. (For example, in systems that have conservation laws, such as conservation of energy or momentum, each value of the conserved variables determines a non-trivial invariant subset of $\Omega$.) However, any non-ergodic system can be split up into "ergodic components"; heuristically, these are minimal invariant subsets of $\Omega$, each of which (except possibly a set of measure zero) supports its own ergodic probability function (this is called the Ergodic Decomposition Theorem; see, e.g., Glasner 2003, p. 72, Theorem 3.22). If we ourselves are part of the system, then we are already confined to one such component. Furthermore, even if the system as a whole is not ergodic, many of its factor systems may be ergodic (see Appendix A). This suggests that, by choosing the right level of description for the system (e.g., by adopting a sufficiently coarse-grained, higher-level description, as discussed in List and Pivato 2015), we may be able to reap the benefits of ergodicity. For the applicability of ergodic methods to non-ergodic Hamiltonian systems, see Section 4 of Berkovitz, Frigg, and Kronz (2006).

This question is central to the entire scientific enterprise. Why are we justified in assuming that scientific laws are the same in different spatial locations, or that they will be the same from one day to the next? Why should replicability of other scientists' experimental results be considered the norm, rather than a miraculous exception? Why is it normally safe to assume that the outcomes of experiments will be insensitive to irrelevant details such as the height of the laboratory bench, or the orientation of the apparatus relative to the planet Jupiter? Why, for that matter, are we justified in the inductive generalizations that are ubiquitous in everyday reasoning?

In effect, we are assuming that the scientific phenomena under investigation are invariant under certain symmetries – both temporal, as discussed earlier, and spatial, as discussed later, including translations, rotations, and so on. But where do we get this assumption from? The answer lies in the principle of Occam's Razor.

Roughly speaking, this principle says that, if two theories are equally consistent with the empirical data, we should prefer the simpler theory.[25] In the present framework, the hypothesis of a symmetry-rich system is simpler than the hypothesis of a symmetry-poor system, other things being equal. The following provisional formulation of the principle of Occam's Razor captures this idea:

> **Occam's Razor:** Given any body of empirical evidence about a temporally evolving system, always assume that the system has the largest possible set of symmetries consistent with that evidence.

We must now make this more precise. We begin by explaining what it means for a particular symmetry to be "consistent" with a body of empirical evidence. Formally, our total body of evidence can be represented as a subset $E$ of $\mathcal{H}$, i.e., namely the set of all logically possible histories that are not ruled out by that evidence. Note that we cannot assume that our evidence is a subset of $\Omega$; when we scientifically investigate a system, we do not normally know what $\Omega$ is. Hence we can only assume that $E$ is a subset of the larger set $\mathcal{H}$ of logically possible histories.

Now let $\psi$ be a transformation of $\mathcal{H}$, and suppose that we are testing the hypothesis that $\psi$ is a symmetry of the system. For any positive integer $n$, let $\psi^n$ be the transformation obtained by applying $\psi$ repeatedly, $n$ times in a row. For example, if $\psi$ is a rotation about some axis by angle $\theta$, then $\psi^n$ is the rotation by the angle $n\theta$.[26] For any such transformation $\psi^n$, we write $\psi^{-n}(E)$ to denote the inverse image in $\mathcal{H}$ of $E$ under $\psi^n$. We say that the transformation $\psi$ is *consistent* with the evidence $E$ if the intersection

$$E \cap \psi^{-1}(E) \cap \psi^{-2}(E) \cap \psi^{-3}(E) \cap \ldots$$

is non-empty. This means that the available evidence (i.e., $E$) does not falsify the hypothesis that $\psi$ is a symmetry of the system.

---

[25] The literature contains many proposals on how to formalize Occam's Razor precisely. See, e.g., Baker (2013) and Fitzpatrick (2015). For an efficiency argument for Occam's Razor, see Kelly (2007).
[26] In the present terms, rotations must be represented as transformations of the state space $X$. In Section 3.5, we represent rotations more explicitly, relying on a formal representation of space.

For example, suppose we are interested in whether cosmic microwave background radiation is *isotropic*, i.e., the same in every direction. Suppose we measure a background radiation level of $x_1$ when we point the telescope in direction $d_1$, and a radiation level of $x_2$ when we point it in direction $d_2$. Call these events $E_1$ and $E_2$. Thus, our experimental evidence is summarized by the event $E = E_1 \cap E_2$. Let $\psi$ be a spatial rotation that rotates $d_1$ to $d_2$. Then, focusing for simplicity just on the first two terms of the infinite intersection above,

$$E \cap \psi^{-1}(E) = E_1 \cap E_2 \cap \psi^{-1}(E_1) \cap \psi^{-1}(E_2).$$

If $x_1 = x_2$, we have $E_1 = \psi^{-1}(E_2)$, and the expression for $E \cap \psi^{-1}(E)$ simplifies to $E_1 \cap E_2 \cap \psi^{-1}(E_1)$, which has at least a chance of being non-empty, meaning that the evidence has not (yet) falsified isotropy. But if $x_1 \neq x_2$, then $E_1$ and $\psi^{-1}(E_2)$ are disjoint. In that case, the intersection $E \cap \psi^{-1}(E)$ is empty, and the evidence is inconsistent with isotropy. As it happens, we know from recent astronomy that $x_1 \neq x_2$ in some cases, so cosmic microwave background radiation is not isotropic, and $\psi$ is not a symmetry.

Our version of Occam's Razor now says that we should postulate as symmetries of our system a maximal monoid of transformations consistent with our evidence. Formally, a monoid $\Psi$ of transformations (where each $\psi$ in $\Psi$ is a function from $\mathcal{H}$ into itself) is *consistent* with evidence $E$ if the intersection

$$\bigcap_{\psi \in \Psi} \psi^{-1}(E)$$

is non-empty. This is the generalization of the infinite intersection that appeared in our definition of an individual transformation's consistency with the evidence. Further, a monoid $\Psi$ that is consistent with $E$ is *maximal* if no proper superset of $\Psi$ forms a monoid that is also consistent with $E$.

> **Occam's Razor (formal):** Given any body $E$ of empirical evidence about a temporally evolving system, always assume that the set of symmetries of the system is a maximal monoid $\Psi$ consistent with $E$.

What is the significance of this principle? Recall that we earlier defined $\Gamma$ to be the set of *all* symmetries of our temporally evolving system. In practice, we do not know $\Gamma$. A monoid $\Psi$ that passes the test of Occam's Razor, however, can be viewed as our best guess as to what $\Gamma$ is.

Furthermore, if $\Psi$ is this monoid, and $E$ is our body of evidence, the intersection

$$\bigcap_{\psi \in \Psi} \psi^{-1}(E)$$

can be viewed as our best guess as to what the set of nomologically possible histories is. It consists of all those histories among the logically possible ones that are not ruled out by the postulated symmetry monoid $\Psi$ and the observed evidence $E$. We thus call this intersection our *nomological hypothesis* and label it $\Omega(\Psi, E)$.

To see that this construction is not completely far-fetched, note that, under certain conditions, our nomological hypothesis does indeed reflect the truth about nomological possibility. If the hypothesized symmetry monoid $\Psi$ is a subset of the true symmetry monoid $\Gamma$ of our temporally evolving system – i.e., we have postulated some of the right symmetries – then the true set $\Omega$ of all nomologically possible histories will be a subset of $\Omega(\Psi,E)$. So, our nomological hypothesis will be consistent with the truth and will, at most, be logically weaker than the truth.

Given the hypothesized symmetry monoid $\Psi$, we can then assume provisionally (i) that any empirical observation we make, corresponding to some event $D$, can be generalized to a $\Psi$-invariant law, as explained in Section 2.6;[27] and (ii) that unconditional and conditional probabilities can be estimated from empirical frequency data using a suitable version of the Ergodic Theorem, as explained in Sections 2.7 and 3.7.

*2.9 Inferential modesty, informational parsimony, and the nomological hypothesis*

Occam's Razor requires us to assume that the symmetries of our system are given by a maximal monoid of transformations consistent with the evidence $E$. Under natural assumptions, at least one such maximal consistent monoid will indeed exist.[28] However, there may be more than one. In this case, we need a criterion to choose one maximal symmetry monoid rather than another. We now develop such a criterion.

Let us begin with an example. Consider a very simple temporally evolving system, where the set $T$ of times contains only a single element. So, histories can be identified with states at that single time; this expositional simplification has no substantive consequences. Suppose that the state of the system is described by a two-dimensional grid of zeros and ones, which is infinite in every direction. Let $X$ be the set of all logically possible grids of this kind. Then the set $\mathcal{H}$ of all logically possible histories can be identified with $X$. In this system, one elementary kind of nomological constraint is one that constrains the values of one or more cells, for example the constraint "In any possible history, the cell (2,3) must have the value zero".[29] Suppose we have obtained evidence that any possible history must satisfy the constraints shown in Figure 1. This evidence would be represented by the subset $E$ of $\mathcal{H}$ consisting of all single-period histories in which the grid coincides with Figure 1 in all non-empty cells.

---

[27] Formally, the $\Psi$-invariant law in question is the intersection, over all $\psi$ in $\Psi$, of $\psi^{-1}(D)$.
[28] For example, if (i) $\mathcal{H}$ has a topology, (ii) $E$ is a compact subset of $\mathcal{H}$, and (iii) all the transformations in question are continuous, then Zorn's lemma implies the existence of a maximal consistent monoid.
[29] Formally, this constraint corresponds to the set $E = \{h \in \mathcal{H} : h(2,3) = 0\}$. Of course, we have chosen this rather artificial example only for expositional simplicity. Typically, we would be interested not so much in nomological constraints on single coordinates, but in constraints on the relationships between two or more coordinates, such as the constraint "No two adjacent cells can both contain a zero".

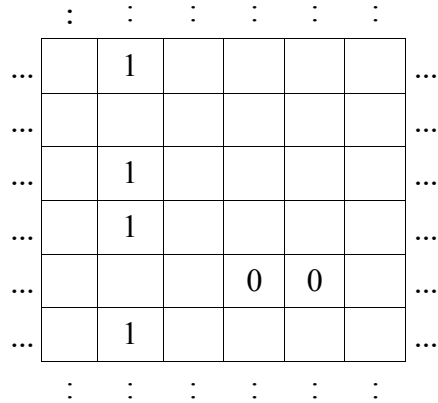

**Figure 1**

Now, for any integer *n*, let $\psi_n^{\rightarrow}$ be the transformation that shifts the entire grid to the *right* by *n* spaces.[30] Let $\Psi^{\rightarrow} := \{ \ldots, \psi_{-1}^{\rightarrow}, \psi_0^{\rightarrow}, \psi_1^{\rightarrow}, \psi_2^{\rightarrow}, \ldots \}$ denote the monoid of all such horizontal shifts. Meanwhile, let $\psi_n^{\uparrow}$ be the transformation that shifts the entire grid *upwards* by *n* spaces, and let $\Psi^{\uparrow} := \{ \ldots, \psi_{-1}^{\uparrow}, \psi_0^{\uparrow}, \psi_1^{\uparrow}, \psi_2^{\uparrow}, \ldots \}$ denote the monoid of all such vertical shifts. Consider two hypotheses:

**Hypothesis 1:** All transformations in $\Psi^{\rightarrow}$ are symmetries of the system.

**Hypothesis 2:** All transformations in $\Psi^{\uparrow}$ are symmetries of the system.

Note that the evidence represented in Figure 1 is consistent with either of these hypotheses. However, it cannot accommodate both of them simultaneously. If Hypothesis 1 were true, then the evidence represented in Figure 1 would entail the constraints shown in Figure 2. By contrast, if Hypothesis 2 were true, then the evidence would entail the constraints shown in Figure 3. In each figure, the constraints that were part of the initial evidence are highlighted in boldface; extrapolated constraints (based on the postulated symmetries) appear in normal, non-bold font. Clearly, Hypotheses 1 and 2 cannot *both* be true, since they yield mutually contradictory constraints on the values of the grey cells.

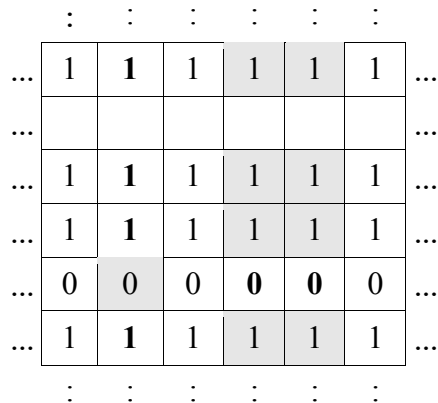

**Figure 2**

---

[30] Of course, if *n* is *negative*, then $\psi_n^{\rightarrow}$ is actually a shift to the *left*.

|   | : | : | : | : | : | : |   |
|---|---|---|---|---|---|---|---|
|...|   | **1** |   | 0 | 0 |   |...|
|...|   | 1 |   | 0 | 0 |   |...|
|...|   | 1 |   | 0 | 0 |   |...|
|...|   | 1 |   | 0 | 0 |   |...|
|...|   | 1 |   | **0** | **0** |   |...|
|...|   | 1 |   | 0 | 0 |   |...|
|   | : | : | : | : | : | : |   |

**Figure 3**

Let $\Psi$ be some maximal consistent monoid of transformations that we postulate as the symmetry monoid, in accordance with Occam's Razor. Hypothesis 1 then asserts that $\Psi^{\rightarrow} \subseteq \Psi$, while Hypothesis 2 asserts that $\Psi^{\uparrow} \subseteq \Psi$. Since both hypotheses cannot simultaneously be true, it follows that there are at least two distinct ways in which we could specify $\Psi$: one including $\Psi^{\rightarrow}$ and another including $\Psi^{\uparrow}$. So even in this very simple example, there is no unique maximal consistent monoid.

At first sight, the choice between these two maximal symmetry monoids seems arbitrary. But it is not. To see this, note that both hypotheses could have entailed the same constraints they did, using less initial evidence. For example, Hypothesis 1 would have entailed the same constraints from the evidence represented in Figure 4.

|   | : | : | : | : | : | : |   |
|---|---|---|---|---|---|---|---|
|...|   | 1 |   |   |   |   |...|
|...|   |   |   |   |   |   |...|
|...|   | 1 |   |   |   |   |...|
|...|   | 1 |   |   |   |   |...|
|...|   |   |   | 0 |   |   |...|
|...|   | 1 |   |   |   |   |...|
|   | : | : | : | : | : | : |   |

**Figure 4**

The original evidence in Figure 1 constrained *six* cell values (i.e., six "bits" of information). But Hypothesis 1 can make do with only *five* of them (in particular, the second zero is redundant). Meanwhile, Hypothesis 2 would have entailed the same constraints from only *three* bits of information, as represented in Figure 5.

|   | : | : | : | : | : | : |   |
|---|---|---|---|---|---|---|---|
| ... |   |   |   |   |   |   | ... |
| ... |   |   |   |   |   |   | ... |
| ... |   |   |   |   |   |   | ... |
| ... |   |   |   |   |   |   | ... |
| ... |   |   |   |   | 0 | 0 | ... |
| ... |   |   | 1 |   |   |   | ... |
|   | : | : | : | : | : | : |   |

**Figure 5**

In other words, Hypothesis 2 could have entailed all of its original constraints, using *less information* than Hypotheses 1 needed to obtain its original constraints. Thus Hypothesis 2 can be viewed as more *informationally parsimonious* than Hypothesis 1. Hypothesis 2 stands out in another way too: from the same initial evidence, it constrains *fewer cell values* than Hypothesis 1. So, Hypothesis 2 is also more *inferentially modest* than Hypothesis 1.

This simple example illustrates two general points. First, different symmetry monoids may lead to different nomological hypotheses – hypotheses about what the nomologically possible histories are – even starting from the same evidence. Formally, we may have $\Omega(\Psi_1,E) \neq \Omega(\Psi_2,E)$ where $\Psi_1$ and $\Psi_2$ are two distinct symmetry monoids that are each consistent with evidence $E$. Second, one symmetry monoid could generate the *same* nomological hypothesis from two different bodies of evidence. Formally, we may have $\Omega(\Psi,E_1) \neq \Omega(\Psi,E_2)$ for the same symmetry monoid $\Psi$ and two distinct bodies of evidence $E_1$ and $E_2$.

Thus, given two symmetry monoids $\Psi_1$ and $\Psi_2$, which are each compatible with the same body of evidence $E$, we can compare them along two dimensions:

> **Inferential modesty:** If $\Omega(\Psi_2,E) \subseteq \Omega(\Psi_1,E)$, then we say that $\Psi_1$ is (at least weakly) more *inferentially modest* than $\Psi_2$.
>
> **Informational Parsimony:** Let $E_1$ be the largest superset[31] of $E$ such that $\Omega(\Psi_1,E_1) = \Omega(\Psi_1,E)$. Let $E_2$ be the largest superset of $E$ such that $\Omega(\Psi_2,E_2) = \Omega(\Psi_2,E)$. If $E_2 \subseteq E_1$, then we say that $\Psi_1$ is (at least weakly) more *informationally parsimonious* than $\Psi_2$.

Returning to our earlier example with the infinite grid, let $E$ be the evidence described by Figure 1. Then $\Omega(\Psi^\rightarrow,E)$ is the set of single-period histories satisfying the constraints described by Figure 2, and $\Omega(\Psi^\uparrow,E)$ is the corresponding set for Figure 3. Meanwhile, if $E^\rightarrow$ is the evidence described by Figure 4, then we have $\Omega(\Psi^\rightarrow,E^\rightarrow) = \Omega(\Psi^\rightarrow,E)$. Likewise, if $E^\uparrow$ is the evidence described by Figure 5, then we have $\Omega(\Psi^\uparrow,E^\uparrow) = \Omega(\Psi^\uparrow,E)$.

---

[31] Recall that larger subsets of $\mathcal{H}$ encode *less* information. In particular, if $E_1$ is a superset of $E$, then $E_1$ encodes *less information* than $E$.

In this example, neither $\Omega(\Psi^{\rightarrow},E)$ nor $\Omega(\Psi^{\uparrow},E)$ includes the other, so neither monoid is more inferentially modest than the other, according to our definition. Likewise, neither $E^{\rightarrow}$ nor $E^{\uparrow}$ includes the other, so neither monoid is more informationally parsimonious. So our formal definitions up to this point are not sensitive enough to capture the plausible intuition that $\Psi^{\uparrow}$ is both more inferentially modest and more informationally parsimonious than $\Psi^{\rightarrow}$.

One possible way of capturing this intuition – though this is not central to this paper – is to use concepts from information theory, such as entropy. To do this, we must introduce a prior probability distribution $Pr_0$ on the set $\mathcal{H}$ of all possible histories. In the example with the infinite grid, this could be the *uniform Bernoulli* distribution, which treats all the cells in the grid as independent, identically distributed random variables, where zero and one each appear with probability ½. Given two different symmetry monoids $\Psi_1$ and $\Psi_2$ that are compatible with the same body of evidence $E$, we can use $Pr_0$ to compare them as follows:

> **Inferential modesty (relative to $Pr_0$):** If $Pr_0[\Omega(\Psi_2,E)] \leq Pr_0[\Omega(\Psi_1,E)]$, then we say that $\Psi_1$ is (at least weakly) more *inferentially modest* than $\Psi_2$, relative to $Pr_0$.

> **Informational Parsimony (relative to $Pr_0$):** Let $E_1$ be the largest superset of $E$ such that $\Omega(\Psi_1,E_1) = \Omega(\Psi_1,E)$. Let $E_2$ be the largest superset of $E$ such that $\Omega(\Psi_2,E_2) = \Omega(\Psi_2,E)$. If $Pr_0[E_2] \leq Pr_0[E_1]$, then we say that $\Psi_1$ is (at least weakly) more *informationally parsimonious* than $\Psi_2$, relative to $Pr_0$.

Do these criteria enable us to prefer $\Psi^{\uparrow}$ to $\Psi^{\rightarrow}$, as intuition suggests? Let us begin with the second criterion. Comparing Figures 4 and 5, we see that $Pr_0[E^{\rightarrow}] = 2^{-5}$, whereas $Pr_0[E^{\uparrow}] = 2^{-3}$, and so $\Psi^{\uparrow}$ is indeed more informationally parsimonious than $\Psi^{\rightarrow}$, relative to the uniform Bernoulli distribution. The first criterion, by contrast, does not help. Comparing Figures 2 and 3, we see that $Pr_0[\Omega(\Psi^{\rightarrow},E)]$ and $Pr_0[\Omega(\Psi^{\uparrow},E)]$ are each zero, because they constrain an infinite number of cells. So, they do not differ in inferential modesty relative to $Pr_0$. They do differ in more sensitive measures of inferential modesty, computed using more advanced notions from information theory, such as "entropy density". But the details are beyond the scope of this paper.

Note that, if $\Psi_1$ is more inferentially modest than $\Psi_2$ in the original sense, which did not refer to any prior probability, then $\Psi_1$ is more inferentially modest than $\Psi_2$ in the information-theoretic sense, relative to *any* prior $Pr_0$. This is because if $\Omega(\Psi_2,E) \subseteq \Omega(\Psi_1,E)$, then $Pr_0[\Omega(\Psi_2,E)] \leq Pr_0[\Omega(\Psi_1,E)]$. Likewise, if $\Psi_1$ is more informationally parsimonious than $\Psi_2$ in the original sense, then $\Psi_1$ is more informationally parsimonious than $\Psi_2$ in the new sense, relative to *any* prior $Pr_0$. The reason is that if $E_2 \subseteq E_1$, then $Pr_0[E_2] \leq Pr_0[E_1]$.

*2.10 The role of time*

What is the significance of the linear order of the set $T$ of times? Why is time ordered in one way, and not in another? Do the laws of a given system "care" about the ordering of time? To put it another way: what does it mean to say that today comes between yesterday and tomorrow? Intuitively, it means this: the events that happened yesterday cannot "directly influence" the events that will happen tomorrow; their influence must be "mediated" by the events that happen today. We now make this claim precise using a standard notion from probability theory: the *Markov property*.[32]

To explain this property, we first introduce the notion of *conditional independence*. Let $\{Pr_E\}_{E \subseteq \Omega}$ be a conditional probability structure, and let $D$ and $E$ be two events (i.e., subsets of $\Omega$). We say that $D$ and $E$ are *independent* if $Pr_D(E) = Pr_\Omega(E)$ and $Pr_E(D) = Pr_\Omega(D)$.[33] Informally, if we interpret probabilities as encoding "information", this means that learning whether or not $D$ has occurred provides no information about whether or not $E$ will occur, and vice versa.

To illustrate, recall the simple coin-tossing system from Section 2.7. Let $E$ and $D$ be the events "the outcome at time 1 is *Heads*" and "the outcome at time 2 is *Tails*". Then $Pr_\Omega(E) = ½$ and $Pr_\Omega(D) = ½$, assuming for simplicity that $p = 0.5$. Here, the outcome at time 1 has no effect on the outcome at time 2. So, even if we tossed *Heads* at time 1, this would not change the probability of obtaining *Tails* at time 2, and so $Pr_E(D) = ½$. Likewise, the outcome at time 2 tells us nothing about what happened at time 1. If we had not observed the outcome at time 1 but obtained the outcome *Tails* at time 2, we would still assign probability ½ to *Heads* at time 1. So, $Pr_D(E) = ½$. Thus, the events $E$ and $D$ are independent.

Now let $C$, $D$, and $E$ be three events. We say that $C$ and $E$ are *conditionally independent*, given $D$, if $Pr_{C \cap D}(E) = Pr_D(E)$ and $Pr_{E \cap D}(C) = Pr_D(C)$. Again, if we interpret probabilities as encoding "information", this means the following. Suppose you already know that $D$ has occurred. Then learning whether or not $C$ has occurred provides no further information about whether or not $E$ will occur, and vice versa.

To illustrate, return again to the coin-tossing example (where $T = \{1,2,3,....\}$) with $p = 0.5$, but suppose that we use the tosses of the fair coin to determine the position of a token on an infinite line. We move the token after each coin toss: if we toss *Heads*, we move the token one space to the right, and if we toss *Tails*, we move it one space to the left. Let us represent the position of the token by an integer (either positive or negative); in other words, $X = \{...,-3,-2,-1,0,1,2,3,...\}$. Let $x_t$ denote the position of the token at time $t$. Then the rule becomes the following: "if you toss *Heads* at time $t$, then $x_{t+1} = x_t + 1$. If you toss *Tails* at time $t$, then $x_{t+1} = x_t - 1$". For simplicity, suppose the coin always starts at position 0 (i.e., $x_1 = 0$).[34]

If $D$ is an event describing the position of the token at time $t$, and $E$ is an event describing its position at time $t+1$, then these two events are not independent. For example, suppose $E$ is the event, "$x_6 = 3$". Then a simple calculation shows that $Pr_\Omega(E) = 5/32$. However, suppose $D$ is the event "$x_5 = 2$". Then $Pr_D(E) = ½$ (because the token now has a 50% probability of moving from position 2 to position 3 in one time step). Thus, *E* and *D* are *not* independent. The location of the token at time 5 tells us a great deal about its probable location at time 6.

---

[32] The importance of Markov properties in understanding causality has been emphasized by Pearl (2000) and Spirtes, Glymour, and Scheines (2000).
[33] If $Pr_\Omega(D) > 0$, the first equation is equivalent to $Pr_\Omega(E \cap D) = Pr_\Omega(D) Pr_\Omega(E)$. If $Pr_\Omega(E) > 0$, the second equation is equivalent to $Pr_\Omega(E \cap D) = Pr_\Omega(D) Pr_\Omega(E)$. Thus, if $Pr_\Omega(D) > 0$ and $Pr_\Omega(E) > 0$, the two equations are equivalent. But if $Pr_\Omega(D) = 0$ or $Pr_\Omega(E) = 0$, the equations must be stated separately.
[34] Technically, the system just described is a *simple random walk*.

However, once we know the position at time 5, learning the position at time 4 tells us *nothing further* about the position at time 6. Continuing the previous example, let *C* be the event "$x_4 = 1$". Then straightforward calculations show that $Pr_{C \cap D}(E) = ½ = Pr_D(E)$ and $Pr_{E \cap D}(C) = ½ = Pr_D(C)$. In other words, if we *already knew* that the token's position was 2 at time 5 (so that it had a 50% probability of moving to position 3 at time 6), then learning its position at time 4 tells us *nothing further* about where it might be at time 6. Likewise, if we *already knew* that the token's position was 2 at time 5 (so that it has a 50% probability of having been at position 1 at time 4), then learning its position at time 6 tells us *nothing further* about where it might have been at time 4.

In this example, the conditional independence of the events *C* and *E*, given *D*, is due to the fact that *D* concerns the state of the system at a point in time *between* the times described by *C* and *E* and that *D* provides us with *complete* information about the state of the system at this intermediate time. If *D* provided only partial information about that state, we would not get the same result. For example, suppose *D'* is the event, "$x_5 = 0, 2,$ or $4$", which does not fully specify the state at time 5. Then it can be shown that $Pr_{C \cap D'}(E) > Pr_{D'}(E)$. Here, learning additional information about the state at time 4 can still tell us something about where the coin is likely to be at time 6.

Now let us generalize this example. Let *T* be any linearly ordered set, let *X* be any set of states, and consider a temporally evolving system given by a collection Ω of possible histories (i.e., functions from *T* into *X*) and a conditional probability structure $\{Pr_E\}_{E \subseteq \Omega}$. For any time *t* in *T*, and any state *x* in *X*, let $E_x^t$ denote the event, "the state of the system at time *t* is *x*". More generally, for any subset *Y* of *X*, let $E_Y^t$ denote the event, "the state of the system at time *t* is an element of *Y*". We say that the system satisfies the *Markov property* if, for any times $r < s < t$ in *T*, any subsets *Y* and *Z* of *X*, and any state *x* in *X*, the events $E_Y^r$ and $E_Z^t$ are conditionally independent, given the event $E_x^s$. In other words, if you have *complete information* about the state of the system at some time *s* (you know that the state is *x*), then learning something about its state at some earlier time (e.g., that it was an element of *Y* at time *r*) tells you *nothing further* about its probable state at some later time (e.g., about how probable it is that it will fall into the set *Z* at time *t*). Roughly speaking, this means that the state of the system at time *r* cannot "directly influence" the state of the system at time *t*. It can only influence that state "indirectly", via influencing the state at the intermediate time *s*. Any system with this property is called *Markovian*.

Note that the Markov property does not say that the future evolution of the system is unconditionally independent of its past. It just says that the dependency of the future upon the past is mediated through the present. This property is fundamental to the way we normally think about time. To see this, imagine a universe where the Markov property was *not* true. Then there would exist some times $r < s < t$ in *T*, some subsets *Y* and *Z* of *X*, and some state *x* in *X*, such that the conditional probability $Pr(E_Z^t \mid E_Y^r \cap E_x^s)$ is not the same as $Pr(E_Z^t \mid E_x^s)$.[35] In other words, even with a

---

[35] Here, to avoid cumbersome subscripts, we are using the notation $Pr(A \mid B)$ to denote the conditional probability $Pr_B(A)$.

*complete* specification of the present state *x*, the probability of some *future* event *Z* would depend on whether or not some *past* event *Y* had occurred. This would suggest that the state specification *x* does *not*, in fact, contain all the information about the present state of the system; somehow, information about the past history is bypassing the present and "leaking" directly into the future. This, in turn, suggests that this so-called "past" is not really in the past at all; our model of the time structure of the system is incorrect.

We take the Markov property to be a necessary condition for the "correct" ordering of time. To be "well-behaved", a temporally evolving system must be Markovian. What the present must do at any point in time in order to *count as* the present is "separate" the past from the future. If this property is violated, the set *T* does not properly play the role of time.

Two points are worth noting. First, some systems may admit multiple time orderings with respect to which they are Markovian. An extreme limiting case is given by our original coin-tossing system without the moving token, which is Markovian with respect to *every* ordering of *T*. Here, the precise order of time is irrelevant. By contrast, in the modified coin-tossing system with the token, the order of time matters, as we have seen. In fact, the temporal order with respect to which the system satisfies the Markov property is essentially unique; it is unique up to time reversals. This brings us to our second point. Although the Markov property says something about the linear "topology" of time, it tells us nothing about the *direction* of time. As illustrated by the modified coin-tossing system, the Markov property is completely invariant under time reversals. In other words, the Markov property only says that the present separates the past from the future. But it does not tell us on which side of the present lies the past, and on which side lies the future.

What, then, can we say about the directionality of time? Using the concepts introduced earlier, we can say that a condition for time to have a direction in a system is that time reversals are *not* symmetries of the system. Since time reversals are symmetries of classical mechanical systems (in the sense explained in footnote 17), it follows that, in those systems, there is no real direction of time: temporal orders are unique at most up to time reversal. By contrast, in thermodynamic systems, time reversals are not symmetries, and hence these systems meet the condition for time to have a direction. To the extent that the world, as seen from our perspective, is best understood as a system in which time reversals are not symmetries, there is then a coherent basis for the directionality of time (for further discussion, see Roberts 2013).

## 3. Spatially extended systems

*3.1 Basic definitions*

We now turn to a more richly described class of systems whose state evolves over time. As before, we represent *time* by a linearly ordered set *T*. We now also incorporate an explicit notion of *space*, represented by a set *S* of *spatial locations*. Let $S \times T$ be the set of all ordered pairs of the form (*s*, *t*), where *s* is an element of *S*, and *t* is an element of *T*. We refer to $S \times T$ as *space-time*. Again, let *X* denote a set of possible states, called the *state space*. Unlike in our earlier model, the elements of *X* are no longer "global" states, in which the system can be *at specific points in time*, but "local" states, in which the system can be *at specific points in space and time*. Again, we treat the elements of *X* as primitives of our model. Histories are now functions from space-time (rather than merely time) into the state space. Formally, a *spatially extended history* is a function *h* from $S \times T$ into *X*. For each point $(s, t)$ in $S \times T$, $h(s, t)$ is the state of the system in spatial location *s* at time *t*.

In analogy to our earlier model, we write $\Omega$ to denote the set of all spatially extended histories deemed *possible*, which, as before, play the role of possible worlds. Again, this is a subset – often a proper one – of the set $\mathcal{H}$ of all *logically possible* such histories (here, all functions from $S \times T$ into *X*). So, membership in $\Omega$ is best interpreted as *nomological possibility*. Subsets of $\Omega$ are called *events*.

Finally, we define a *conditional probability structure* on $\Omega$. As before, this is a family of conditional probability functions $\{Pr_E\}_{E \subseteq \Omega}$, one $Pr_E$ for each event *E* in $\Omega$, satisfying standard consistency conditions. Recall that $Pr_E$ assigns to any event in $\Omega$ the conditional probability of that event, given *E*. A *spatially extended system* is the pair consisting of the set $\Omega$ of possible spatially extended histories and the conditional probability structure $\{Pr_E\}_{E \subseteq \Omega}$.

For example, in a classical mechanical system, *T* is the set **R** of real numbers, *S* is the three-dimensional Euclidean space (formally, $S = \mathbf{R}^3$), and each state $h(s,t)$ in *X* is given by the set of particles present at spatial location *s* at time *t*, along with their physically relevant properties (e.g., masses and momenta) and the values of any force fields (e.g., gravity) acting on these particles.[36] In a classical electrodynamical system, the state $h(s, t)$ must also specify the particles' charges, along with the electric and magnetic field vectors at $(s, t)$. In that sense, electrodynamics relies on a richer ontology than classical mechanics.

In a quantum-mechanical system, it might be tempting to suppose that $S = \mathbf{R}^3$, and to suppose that $h(s, t)$ is given by the values of the wave functions of each of the particles in the system at space-time location $(s, t)$. But this is not correct, because the wave functions of interacting particles in a quantum system cannot generally be defined independently of each other. Instead, we must define a *joint* wave function for the entire multi-particle system. So, in a quantum-mechanical system with *n* particles, we would define space to be $S = (\mathbf{R}^3)^n$, with three coordinates representing the spatial "position" of each of the *n* particles in an underlying ordinary Euclidean space;[37] and we would define the set *X* of possible states of the system to be the set of complex numbers, capturing amplitudes, whose squared absolute values behave formally like probabilities. Thus a spatially extended history *h* is a function from $(\mathbf{R}^3)^n \times T$ into the set of complex numbers, representing the joint wave function of the whole ensemble of particles.

For instance, if there are two particles, labeled 1 and 2, then $h(x_1, y_1, z_1, x_2, y_2, z_2, t)$ represents the joint state at time *t* of particles 1 and 2 at positions $x_1, y_1, z_1$ and $x_2, y_2, z_2$ in the underlying three-dimensional Euclidean space. This joint state of the two particles is a complex number whose squared absolute value can be interpreted, under some assumptions, as the probability of particles 1 and 2 being observable at positions $x_1, y_1, z_1$ and $x_2, y_2, z_2$, respectively, at time $t$.

---

[36] We are *not* saying that this is the most parsimonious or computationally convenient way to represent a classical mechanical system. It is only one way of representing such a system in our framework.

[37] Strictly speaking, particles in quantum systems do not have "positions", so we are using this term rather loosely. Also, there is a dual representation of the wave function (obtained via Fourier transform), where the coordinates in $(\mathbf{R}^3)^n$ represent the "momenta" (again, loosely) of the *n* particles. These two representations are equally valid.

*3.2 Determinism and indeterminism*

As with our original, simple class of temporally evolving systems, we can define different notions of determinism and indeterminism for spatially extended systems. For any subset $L$ of locations in $S \times T$, we write $h_L$ to denote the restriction of the function $h$ to the points in $L$. We can then ask for which proper subsets $L$ of $S \times T$, if any, it is true that $h_L$ has a unique extension to all of $S \times T$ in $\Omega$. Again, an *extension* of $h_L$ is a history $h'$ such that $h'_L = h_L$. When $h_L$ is uniquely extendible to all of $S \times T$, we say that the history $h$ is *L-deterministic*.

For example, the histories of classical mechanical systems are *L*-deterministic for any subset $L$ of $S \times T$ that has the form $S \times T'$, where $T'$ is any non-empty subset of $T$. Information about the system for even a single "time slice" of space-time, i.e., a set of the form $S \times \{t\}$ for some $t$ in $T$, is sufficient to determine the full spatially extended history. In contrast, the histories of quantum-mechanical systems (assuming the possibility of wave-function collapses) are not generally *L*-deterministic for time slices or collections of time slices.

The present definitions allow us to explore some interesting possibilities not captured by standard definitions that focus exclusively on past-to-future determination. For example, some systems might encode their entire spatially extended history in each individual space-time location. Histories would then be *L*-deterministic for every singleton set $L=\{(s,t)\}$, where $(s,t)$ is in $S \times T$. Here, we would have an extreme form of local-to-global determinism. Alternatively, some systems might encode their entire spatially extended history in some collection of "spatial slices of time", i.e., some subset $L$ of $S \times T$ which has the form $S' \times T$, where $S'$ is a non-empty subset of $S$, possibly singleton. This would be a kind of spatial, rather than temporal, determinism.[38] Other systems might never be *L*-deterministic for any proper subset $L$ of $S \times T$.

There may also be some intermediate forms of determination, for instance when a history restricted to some set $L$ of locations is uniquely extendible to a history restricted to some superset $L^*$ of $L$, which is still smaller than $S \times T$ in its entirety.[39] We might imagine, for instance, systems that are deterministic strictly "across space" but not "across time". In such a system, a history restricted to some set $L$ of the form $S' \times \{t\}$, where $S'$ is a non-empty subset of $S$ and $t$ a point in time, might determine the entire "time slice" of that history across $L^* = S \times \{t\}$, but not the rest of the history. Some crystals and other chemical or physical systems involving highly regular spatial structures might have this feature. Similarly, for suitable specifications of $L$ and $L^*$, we can represent the phenomenon that, in systems in which "information" travels with finite speed, events at particular space-time locations determine events within their "light cones", but not outside them.

---

[38] This sort of determinism occurs in *expansive cellular automata*, a class of spatially extended systems discussed in the theory of dynamical systems. See Pivato (2009).

[39] For example, this phenomenon arises frequently in the solution of *boundary value problems* in mathematical physics. See Pivato (2010).

*3.3 Nomological possibility and necessity*

In analogy to the case of temporally evolving systems, we can define two modal operators for each set *L* of space-time locations, namely nomological possibility and necessity relative to *L*. For each set $L \subseteq S \times T$, call one history, *h'*, *accessible* from another, *h*, relative to *L*, if the restrictions of *h* and *h'* to *L* coincide, i.e., $h'_L = h_L$. We then write $h'R_Lh$. For any event $E \subseteq \Omega$, we define

◆$_L E$ = {$h \in \Omega$ : for some $h' \in \Omega$ with $h'R_Lh$, we have $h' \in E$},

■$_L E$ = {$h \in \Omega$ : for all $h' \in \Omega$ with $h'R_Lh$, we have $h' \in E$}.

Here, ◆$_L E$ and ■$_L E$ are, respectively, the sets of all histories in which *E* is nomologically possible and nomologically necessary *once the system's history within space-time region L is given*. Important special cases are (i) $L = S \times T'$, where *S* is all of space and *T'* is the set of all points in time up to some time *t*, (ii) $L = S' \times T$, where *T* is all of time and *S'* is some spatial region, and (iii) $L = \varnothing$ for possibility and necessity in the "atemporal" sense. Since the present definitions are completely analogous to their counterparts in Section 2.3 for temporally evolving systems, we will not say more about them here.

*3.4 Constraints and correlations*

We now turn again to the question of how to distinguish between those constraints and correlations that are "brute facts", and those that are genuine "laws" of a system. As in the original case of temporally evolving systems, our analysis is based on the notion of *symmetry*, only now with the additional ingredient that these symmetries can involve space as well as time.

In analogy to our earlier definition, we define a *constraint* to be a property, *C*, that a spatially extended history may or may not have. As before, any constraint *C* can be associated with some subset [*C*] of the set $\mathcal{H}$ of all logically possible histories, where [*C*] is called the *extension* of *C*. A spatially extended history satisfies *C* if it belongs to [*C*]. The spatially extended system as a whole *satisfies C* if *all* histories in $\Omega$ satisfy *C*.

Likewise, as before, any *probabilistic property*, *P*, is associated with its *extension*, which is a subset, denoted [*P*], of the set $\Pi$ of all logically possible conditional probability structures on $\Omega$. A conditional probability structure $\{Pr_E\}_{E \subseteq \Omega}$ satisfies *P* if it belongs to [*P*].

*3.5 Symmetries*

The notion of a *state symmetry* for spatially extended systems is virtually identical to the one defined in Section 2.5 for temporally evolving systems, so we do not discuss it further.[40] Instead, we turn directly to *symmetries acting on space-time*.

---

[40] Technically, given any function ϕ from *X* into itself, and any spatially extended history *h* in $\Omega$, we can define a transformed history ϕ(*h*)=*h'*, where, for all (*s,t*) in $S \times T$, $h'(s,t) = \phi[h(s,t)]$. As before, ϕ is a *state symmetry* if (i) ϕ(*h*) is in $\Omega$, for all *h* in $\Omega$, and (ii) for any events *E'* and *D'* in $\Omega$, if *E* and *D* are

Let ψ be a function from $S \times T$ into itself (i.e., a transformation of space-time). Again, ψ induces a function from the set $\mathcal{H}$ of logically possible histories into itself. For any spatially extended history $h$, we define the transformed history

$$\psi(h) = h', \text{ where, for all } (s,t) \text{ in } S \times T, h'(s,t) = h[\psi(s,t)].$$

As before, given any set $E$ of histories in $\mathcal{H}$, the *inverse image* of $E$ under ψ, written $\psi^{-1}(E)$, is the set of all histories $h$ in $\mathcal{H}$ such that $\psi(h)$ lies in $E$. The function ψ is a *symmetry* if

- $\psi(h)$ is in $\Omega$, for all $h$ in $\Omega$; and
- for any events $E$ and $D$ in $\Omega$, if $E'$ and $D'$ are the inverse images of $E$ and $D$ under ψ, then $Pr_{E'}[D'] = Pr_E[D]$.[41]

For example, if $T$ is the set of real numbers (i.e., $T = \mathbf{R}$), and $S$ is the three-dimensional Euclidean space (i.e., $S = \mathbf{R}^3$), we can consider a spatially extended system in classical mechanics. The following transformations of $S \times T$ are space-time symmetries of such a system, each defined for all $(s,t)$ in $S \times T$:

- *Time translation*: $\psi(s,t) = (s, t+r)$, where $r$ is a fixed real number;
- *Spatial translation*: $\psi(s,t) = (s+v, t)$, where $v$ is a fixed three-dimensional vector (an element of $\mathbf{R}^3$); and
- *Space-time rescaling*: $\psi(s,t) = (r\,s,\, r\,t)$, where $r > 0$ is a fixed real number.

More general symmetries of spatially extended systems include composite functions resulting from the combination of a transformation φ of the state space ($X$) with a transformation ψ of space-time ($S \times T$).[42] Examples in classical mechanical systems are spatial rotations, spatial reflections, spatial rescalings, and Galilean transformations.[43] Crucially, it is possible that neither the transformation φ of the state

---

the inverse images of $E'$ and $D'$ under φ, then $Pr_E[D] = Pr_{E'}[D']$. For example, consider an *n*-particle quantum system, where $S = (\mathbf{R}^3)^n$, $X$ is the set of complex numbers, and a spatially extended history $h$ is a wave function. Let φ be a *phase rotation* map on the complex plane; formally, there is some angle θ such that, for all $x$ in $X$, $\phi(x) = e^{i\theta}x$. Then φ is a state symmetry of the quantum system.

[41] As before, for any subsets $D$, $E$ of $\mathcal{H}$, we define $Pr_E(D) = Pr_{E \cap \Omega}(D \cap \Omega)$.

[42] An additional property we might require of a space-time transformation ψ is *time preservation*: ψ is *time-preserving* if, for any points $(s_1,t_1)$ and $(s_2,t_2)$ in $S \times T$, with $\psi(s_1,t_1) = (s'_1,t'_1)$ and $\psi(s_2,t_2) = (s'_2,t'_2)$, if $t_1 \leq t_2$, then $t'_1 \leq t'_2$. In particular, this implies that if $t_1 = t_2$, then $t'_1 = t'_2$. A time-preserving transformation acts on $S \times T$ such that the set of all space-time points "at time $t_1$" gets moved *en bloc* to the set of all space-time points "at time $t'_1$". All of the transformations described above are time-preserving, but technically, we do not need to include it in our formal definition.

[43] These are defined as follows. *Spatial rotation*: Fix a line $L$ in $S$ and an angle θ. For any point $s$ in $S$, let $s'$ be the point obtained by rotating $s$ by an angle of θ around $L$. For all $(s,t)$ in $S \times T$, define $\psi(s,t) = (s', t)$. Let $L'$ be the line parallel to $L$, but passing through the origin. For all $x$ in $X$, define $\phi(x)$ by rotating all the momentum vectors in $x$ by the angle θ around $L'$. *Spatial reflection*: Fix a plane $P$ in $S$. For any point $s$ in $S$, let $s'$ be the point obtained by reflecting $s$ across $P$. For all $(s,t)$ in $S \times T$, define $\psi(s,t) = (s', t)$. Let $P'$ be the plane parallel to $P$, but passing through the origin. For all $x$ in $X$, define $\phi(x)$ by reflecting all the momentum vectors in $x$ across $P'$. *Spatial rescaling*: Fix some real number $r > 0$, and define $\psi(s,t) = (rs, t)$ for all $(s,t)$ in $S \times T$. Meanwhile, let φ be a transformation of $X$ that multiplies the momentum vector of every particle by $r$, and also multiplies all force field vectors by $r$. *Galilean transformation*: For all $(s,t)$ in $S \times T$, define $\psi(s,t) = (s + t\,v, t)$, where $v$ is a fixed three-

space nor the transformation ψ of space-time alone is a symmetry, and yet, when combined, the two transformations form a symmetry.[44]

Of course, any combination of symmetries is also a symmetry. An example is a *spatiotemporal translation*, which is a combination of a time translation and a spatial translation. In a classical electrodynamical system, *only* the spatiotemporal translations are space-time symmetries. Galilean transformations are not space-time symmetries of classical electrodynamics; indeed, this was the original impetus for the development of special relativity theory.

*3.6 Laws*

Let *C* be a constraint on a spatially extended system, with extension [*C*], and let γ be a symmetry of the system. Recall that *C* is *invariant* under γ if the set [*C*] is equal to its inverse image under γ. Let Γ be the set of all symmetries of the spatially extended system. As in Section 2.6, a constraint *C* that is satisfied by the system is a *law* if it is invariant under *all* symmetries in Γ. Likewise, a probabilistic property *P* that is satisfied by the conditional probability structure $\{Pr_E\}_{E \subseteq \Omega}$ is a *probabilistic law* if it is invariant under all symmetries in Γ.

For example, let $S = \mathbf{R}^3$ and $T = \mathbf{R}$, and suppose that Γ contains all the spatiotemporal translations defined in the previous section. Suppose the system satisfies the constraint *C* which says: "if the state of the system at space-time position (3,7,2,14) is *x*, then at position (4,8,1,17) it is *y*". If ψ is a spatial translation by the vector (1,2,3), then the inverse image of [*C*] under ψ corresponds to the constraint *C'* which says: "if the state of the system at (4,9,5,14) is *x*, then at position (5,10,4,17) it is *y*". Clearly, [*C'*] is not the same as [*C*]. Thus, *C* is *not* invariant under ψ, and so *C* is not a law of the system; it is simply a constraint the system happens to satisfy.

However, suppose *C* is the constraint: "for any location $(s_1, s_2, s_3)$ in *S* and any time *t* in *T*, if the state of the system at space-time position $(s_1, s_2, s_3, t)$ is *x*, then at position $(s_1+1, s_2+1, s_3-1, t+3)$ it is *y*". It is easy to see that [*C*] *is* invariant under all spatiotemporal translations. Suppose that Γ consists *only* of the spatiotemporal translations. Then *C* is invariant under *all* elements of Γ, and so *C* is a law of the system.

An illustration is *Gauss's Law* in an electrodynamical system. This states, roughly, that the net "flux" of the electric field passing through the walls of any closed compartment is proportional to the net charge contained inside that compartment. This constraint is invariant under both spatiotemporal translations and rotations, because both the net flux and the net charge are unchanged by these sorts of

---

dimensional vector (an element of $\mathbf{R}^3$). Meanwhile, for all *x* in *X*, define φ(*x*) by adding the vector *v* to all momentum vectors in *x*.

[44] In all three examples just given, neither ψ nor φ is itself a symmetry of classical mechanics. But when combined, they *do* form a symmetry. For another example, in the setting of classical electrodynamics, let ψ be a spatial reflection acting on *S* × *T*, and let φ be a transformation of *X* which applies the corresponding reflection to all momentum vectors and field vectors, and which furthermore *negates* the magnetic field vector. Neither one of these transformations alone is a symmetry of classical electrodynamics, but when combined together, they *do* form a symmetry.

transformations. Indeed, Gauss's Law is preserved by *every* symmetry of an electrodynamical system; that is why it is a law.

*3.7 From local observations to global laws*

Extending the ideas from Section 2.7, we now discuss how space-time symmetries allow us to infer general laws from specific observations. For example, let $S = \mathbf{R}^3$ and $T = \mathbf{R}$, and let $Y$ and $Z$ be two subsets of the state space $X$. Suppose we have made the observation $O$: "whenever the state is in the set $Y$ at space-time position (3,7,2,14), there is a 50% probability that it will be in the set $Z$ at position (4,8,1,17)". If we can assume that all the spatiotemporal translations defined in Section 3.5 are symmetries of the system, we are able to deduce the following general law: "for any location $(s_1, s_2, s_3)$ in $S$ and any time $t$ in $T$, if the state of the system is in the set $Y$ at space-time position $(s_1, s_2, s_3, t)$, then there is a 50% probability that it will be in the set $Z$ at position $(s_1+1, s_2+1, s_3-1, t+3)$".

Again, however, we may not be able to make many observations of possible histories at space-time position (3,7,2,14), perhaps because we cannot "re-run" history many times, so that we may not really be able to observe $O$. The solution to this problem, as before, lies in the property of *ergodicity*.

Recall that, for some collection $\Psi$ of symmetries, an event $E$ (a subset of $\Omega$) is $\Psi$-*invariant* if the $\psi$-inverse image of $E$ is $E$ itself, for all $\psi$ in $\Psi$. For illustrative purposes, let $\Psi$ be the collection of all spatiotemporal translations, as defined in Section 3.5, and suppose all of them are symmetries of the system.[45] The system is *spatiotemporally ergodic* if, for any $\Psi$-invariant event $E$, we have either $Pr_\Omega(E) = 0$ or $Pr_\Omega(E) = 1$, where $Pr_\Omega(E)$ is the unconditional probability of $E$.

Since $\Psi$ consists of spatiotemporal translations, $\Psi$-invariant events are events one cannot escape from either by travelling through space, or by travelling forwards or backwards through time. Returning to our example, let $\psi$ be a spatiotemporal translation in $\Psi$ such that, for all $(s_1, s_2, s_3)$ in $S$ ($=\mathbf{R}^3$) and $t$ in $T$ ($=\mathbf{R}$), we have $\psi(s_1, s_2, s_3, t) = (s_1+5, s_2-7, s_3+10, t+3)$. If we interpret the spatially extended history $h$ as describing a possible world "from the perspective of position (0,0,0,0)", then the transformed history $\psi(h)$ describes the same world "from the perspective of position (5,–7,10,3)". Here a $\Psi$-invariant event $E$ has the property that whenever a history $h$ is in $E$, then so is $\psi(h)$. Roughly speaking, this means that, if the world described by $h$ appears to be in the set $E$ "from the perspective of position (0,0,0,0)", then it will also appear to be in $E$ "from the perspective of position (5,–7,10,3)", and so on. Ergodicity requires such events to occur either almost always (with probability 1) or almost never (with probability 0).

In a spatiotemporally ergodic system, it is possible to estimate the probability of any event "empirically", by simply counting the spatiotemporal frequency with which that event occurs.

---

[45] Again, the statements in this section hold for any collection of symmetries that forms an *amenable group*.

> **Spatiotemporal Ergodic Theorem:** Suppose the system is spatiotemporally ergodic. Let $E$ be any event and let $h$ be any spatially extended history. For all $r > 0$, let $\Psi_r$ be the set of all spatiotemporal translations by any vector $(v_1, v_2, v_3, v_4)$ with integer coordinates, all of which are between 1 and $r$. Let $N_r$ be the number of translations $\psi$ in $\Psi_r$ such that $\psi(h)$ is in $E$. Then, with probability 1, the ratio $N_r / r^4$ will converge to $Pr_\Omega(E)$ as $r$ increases towards infinity.[46]

Intuitively, $N_r$ is the number of times the event $E$ has "occurred" in the spatially extended history $h$ from time 1 to time $r$ and inside a three-dimensional box with side-length $r$. The ratio $N_r / r^4$ is therefore the *frequency* of occurrence of event $E$, up to time $r$ inside this box, in the spatially extended history $h$. This frequency might be measured, for example, by performing a sequence of experiments or observations inside this box. The Spatiotemporal Ergodic Theorem says that, almost certainly (i.e., with probability 1), the empirical frequency will converge to the true probability of $E$ as the number of observations becomes large.[47] As explained in Section 2.7, we can use this procedure to estimate conditional probabilities and in this way learn the properties of the conditional probability structure $\{Pr_E\}_{E \subseteq \Omega}$. Once again, the Ergodic Theorem justifies generalizations from local observations to global laws.

*3.8 The role of space*

What is the significance of "space" in a spatially extended system? As we will now see, its significance lies in the fact that the structure of space affects the way the system evolves over time. To make this precise, we first introduce a formal representation of the topology of space and then discuss the role that it plays in the system's dynamics.

The topology of space can be represented by a binary relation $\rightarrow$ between subsets of $S$. Heuristically, if $R$ and $R'$ are two subsets of $S$, say two "regions" of space, then $R \rightarrow R'$ means that information from $R$ can flow "directly" into $R'$, without needing to pass through some intervening points "between" $R$ and $R'$. Later, we explain exactly what we mean by "information flow", but for the purposes of our initial heuristic discussion, we leave it unexamined. We refer to $\rightarrow$ as the *adjacency structure* of space.

Adjacency structures arise naturally in many systems. For example, suppose $S$ is ordinary three-dimensional Euclidean space, and suppose information can flow only "continuously" through this space. This would be the case, for instance, in a system consisting of particles travelling along continuous trajectories and interacting via continuous force fields, such as those found in classical mechanics, or in a system described by partial differential equations, such as those found in quantum mechanics, classical electrodynamics, or hydrodynamics. In such systems, for any subsets $R$ and

---

[46] Once again, we have stated this result somewhat informally. For a more formal statement, see Krengel (1985, Chapter 6).

[47] It is not necessary to average over a sequence of "boxes"; the same argument works for any sequence of sets which increase in size and thickness in an appropriate sense, technically any *Følner sequence*.

*R'* of *S*, we have *R*→*R'* if there exists a point *s* in *R* such that, for any radius *r* > 0, the ball of radius *r* centred at *s* intersects *R'*.[48]

For another example, suppose *S* is the *three-dimensional integer lattice*: the set of all ordered triples $s = (s_1, s_2, s_3)$, where $s_1$, $s_2$, and $s_3$ are integers. Say that two points *s* and *s'* in *S* are *neighbours* if they differ in only one coordinate and that difference is 1. Thus (3,7,5) and (3,6,5) are neighbours. Suppose information can flow only directly between neighbours in the lattice. Then, for any subsets *R* and *R'* of *S*, we have *R*→*R'* if some point in *R* is a neighbour of some point in *R'*.[49] Discrete spatial geometries of this kind can be found in a class of systems called *cellular automata*.[50]

For a final example, consider a *directed graph*; this is a mathematical structure consisting of a set of "vertices", along with a set of "arrows" which link together pairs of vertices. Directed graphs can be used to model electric circuits, communication networks (e.g., the internet), economic and transportation networks, and biological systems (e.g., neural networks, gene regulatory networks, and epidemiological networks). Suppose *S* is the set of vertices in this graph. Then, for any subsets *R* and *R'* in *S*, we have *R*→*R'* if there is an arrow from some vertex in *R* to some vertex in *R'*.

If the sets *R* and *R'* overlap (i.e., if $R \cap R' \neq \varnothing$), then clearly we have both *R*→*R'* and *R'*→*R*. However, the examples above show that we can have *R*→*R'* even if *R* and *R'* do not overlap, as long as the two sets "touch" each other in some sense. Heuristically, *R*→*R'* means that it is not possible to interpose any "barrier" between *R* and *R'*; there is no "gap" between them.

What role does the adjacency structure play in our model? Why does space have one adjacency structure rather than another? Just as we argued earlier in the case of time, we will now argue that a "correct" adjacency structure on space is one that satisfies a Markov property with respect to the conditional probability structure $\{Pr_E\}_{E \subseteq \Omega}$. This Markov property is defined by considering conditional probabilities based on "partial information" about a spatially extended history.

We therefore need a precise way to talk about such "partial information". Let *R* be a subset of *S*, and let *R* × *T* be the set of all ordered pairs (*s*,*t*), where *s* is an element of *R*, and *t* is an element of *T*. So, *R* × *T* is the set of all time-slices restricted to the spatial region *R*. For any history *h* in Ω, recall that $h_{R \times T}$ denotes the restriction of *h* to the set *R* × *T*. Heuristically, this restriction records only the part of the history *h* which "happens inside *R*". Let us then define the event $[h_{R \times T}]$ to be the set of all extensions of $h_{R \times T}$ to full histories in Ω, i.e., the set of *all* histories *h'* in Ω such that $h_{R \times T} = h'_{R \times T}$. These are precisely the histories that are accessible from *h* relative to the space-time region *R* × *T*. The Markov property for adjacency structures will be based on conditional independence with respect to such events, in the following way.

For any event *E* (i.e., a subset of Ω), we say that *E happens inside R* if, for all histories *h* and *h'* such that $h_{R \times T} = h'_{R \times T}$, the history *h* is an element of *E* if and only if *h'* is an element of *E*. In other words, the question of whether or not a particular history is an element of $E$ is completely determined by the restriction of that history to spatial "region" $R$.

---

[48] Generally, an adjacency structure can be defined in a similar way on any metric or topological space.
[49] Generally, an adjacency structure can be defined in a similar way on any Cayley graph of any group.
[50] See Ilachinski (2001) and Moore and Mertens (2011).

A *tripartition* of $S$ is a triple $(R, R', R'')$, where $R$, $R'$, and $R''$ are three disjoint subsets of $S$ which together cover $S$ (i.e., $R \cup R' \cup R'' = S$), such that it is *not* the case that $R \rightarrow R''$ or $R'' \rightarrow R$. Heuristically, this means that the set $R'$ "separates" $R$ from $R''$. For example, suppose $S$ is three-dimensional Euclidean space, with the adjacency structure introduced above. Let $R$ be the set of all points whose distance to the origin is less than 1: the *unit ball*. Let $R'$ be the set of all points whose distance to the origin is between 1 and 2, so $R'$ is a sort of thick spherical "shell" around $R$. Finally, let $R''$ be the set of all points whose distance to the origin is greater than 2. Then $(R, R', R'')$ is a tripartition of $S$.

We say that the adjacency structure $\rightarrow$ satisfies the *Markov property* with respect to the conditional probability structure $\{Pr_E\}_{E \subseteq \Omega}$ if, for any tripartition $(R, R', R'')$ and any history $h$ in $\Omega$, any event which happens inside $R$ is conditionally independent from any event which happens inside $R''$, given everything that happens in $R'$ (i.e., given $[h_{R' \times T}]$).

Heuristically, this means that there is no way for information to propagate from $R$ into $R''$, or vice versa, without first passing through $R'$. For example, suppose $S$ is three-dimensional Euclidean space, and $(R, R', R'')$ is the "concentric sphere" tripartition described above. In this case, the spherical shell $R'$ acts as a barrier that isolates the ball-shaped compartment $R$ from any influences coming from the "outer region" $R''$. If we have complete information about the history inside $R'$ (i.e., we know $[h_{R' \times T}]$), then we have complete control over the boundary conditions for any experiment we conduct inside $R$, and thus we do not need to control or even know what happens in the outer region $R''$.

As this example shows, scientists implicitly assume that space satisfies the Markov property every time they construct a laboratory apparatus that "isolates" some experiment from the surrounding environment. Indeed, people also implicitly assume the Markov property every time they close the doors and windows of their houses to keep out the cold. Thus, the Markov property is fundamental to the way we ordinarily think of space. It underpins the adjacency structure of space in the same way it underpins the order structure of time.

Just as with time, however, the Markov property does not *completely* determine the structure of space. First, there may be more than one adjacency structure on $S$ which satisfies the Markov property with respect to $\{Pr_E\}_{E \subseteq \Omega}$, just as there may be more than one Markovian order on $T$. Second, the adjacency structure alone leaves many important geometric properties of $S$ unspecified. For example, in many contexts, we would like to define a *metric* on $S$, which determines a notion of "distance" between points. This is obviously crucial in classical mechanics, for example. The adjacency structure does not determine a unique metric. We therefore now turn to the question of how we might arrive at such a metric.

*3.9 Duration and distance*

Recall that the set $T$ of times is linearly ordered. In many contexts, we would like to define a notion of *duration* on $T$. In other words, given four moments $t_1$, $t_2$, $t_3$, and $t_4$ in $T$, with $t_1 < t_2$ and $t_3 < t_4$, we would like to determine whether the time interval between $t_1$ and $t_2$, is greater or smaller than the interval between $t_3$ and $t_4$. To do this, we suppose that the monoid $\Gamma_T$ of temporal symmetries acts *freely and transitively* on $T$. This means that, for any times $t_1$ and $t_2$ in $T$, there is a *unique* symmetry $\gamma$ in $\Gamma_T$ such that $\gamma(t_1) = t_2$. We can then define a formal "subtraction" operation on $T$ as follows. Fix some reference time $t_0$. Now, for any times $t_1$ and $t_2$ in $T$, we define

$t_2 - t_1 = \gamma(t_2)$, where $\gamma$ is the unique temporal symmetry in $\Gamma_T$ such that $\gamma(t_1) = t_0$.

In particular, this implies that $t - t_0 = t$, for any $t$ in $T$. For any four points $t_1$, $t_2$, $t_3$, and $t_4$ in $T$, we say that the time interval from $t_1$ to $t_2$ is greater than the time interval from $t_3$ to $t_4$ if $t_2 - t_1 > t_4 - t_3$. Similarly, we can define a formal "addition" operation on $T$. For any times $t_1$ and $t_2$ in $T$, we define

$t_1 + t_2 = \gamma(t_2)$, where $\gamma$ is the unique temporal symmetry in $\Gamma_T$ such that $\gamma(t_0) = t_1$.

The set $T$, together with the ordering $<$ and the operation $+$, is a *linearly ordered group*.[51]

In many contexts, we would also like to define a *metric* on $S$, which determines a notion of "distance" between points in space. As we have noted, the adjacency structure does not determine a unique metric. But we can define a concept of distance on $S$ by measuring how long it takes for information to travel from one point to the other. To do this, we need to use the concept of duration we have just introduced.

Given any two regions $R$ and $R'$ of $S$, and a time $t$ in $T$, we define what it means for region $R'$ to be "not reachable" from region $R$ in time $t$. We begin with some preliminary definitions. For any subset $R$ of $S$, and any time $t$ in $T$, let $R \times \{t\}$ denote the set $\{(s,t): s \in R\}$. Adapting our earlier definition, we say that an event $E$ *happens inside $R$ at time $t$* if, for all histories $h$ and $h'$ such that $h_{R\times\{t\}} = h'_{R\times\{t\}}$, the history $h$ is an element of $E$ if and only if $h'$ is an element of $E$. In other words, the question of whether or not a particular history is an element of $E$ is completely determined by the restriction of that history to space-time region $R \times \{t\}$. Further, let $R^C$ denote the *complement* of $R$ in $S$, i.e., $R^C = \{s \in S: s \notin R\}$. Given any two subsets $R$ and $R'$ of $S$, and a time $t$ in $T$ with $t > t_0$, we now say that $R'$ is *not reachable from $R$ in time $t$* if, for any history $h$ in $\Omega$, any event which happens in $R'$ at time $t$ is conditionally independent of any event which happens in $R$ at $t_0$, given $[h_{R^C\times\{t_0\}}]$. Informally, once we have complete information about the state of the system *outside* the set $R$ at time $t_0$, learning something about the state of the system *inside* $R$ at $t_0$ gives us

---

[51] Formally, this means that (i) the operation $+$ is *associative*, i.e., $(t_1+ t_2) + t_3 = t_1 + (t_2+ t_3)$ for all $t_1$, $t_2$, $t_3$ in $T$; (ii) there is an *identity* element, namely $t_0$, such that $t_0 + t = t = t + t_0$ for all $t$ in $T$; (iii) every element $t$ in $T$ has an *inverse* $-t$ such that $t + (-t) = t_0 = (-t) + t$; and (iv) the ordering $<$ is *homogeneous*, meaning that, for all $t_1$, $t_2$, $t_3$ in $T$, we have $(t_1 + t_2 < t_1 + t_3) \Leftrightarrow (t_2 < t_3) \Leftrightarrow (t_2 + t_1 < t_3 + t_1)$. It is *not*, in general, the case that linearly ordered groups are *commutative* ("*abelian*"), i.e., we could have $t_1 + t_2 \neq t_2 + t_1$ for some $t_1$, $t_2$ in $T$. See, e.g., Fuchs (2011) for an introduction to ordered groups.

*no further information* about the eventual state of the system inside $R'$ at the later time $t$.[52]

We now define the *distance* $d(R, R')$ between $R$ and $R'$ to be the *maximum* time $t$ in $T$ such that $R'$ is not reachable from $R$ in time $t$, if this maximum exists.[53] This can be interpreted as the *minimum length of time* required for information to "propagate" from $R$ to $R'$. It would be natural to suppose that this notion of distance satisfies the following three properties:

**Symmetry:** For all subsets $R, R'$ of $S$,

$$d(R, R') = d(R', R).$$

**Triangle Inequality:** For all subsets $R, R', R''$ of $S$,

$$d(R, R'') \leq d(R, R') + d(R', R'').$$

**Non-Complementarity:** For all subsets $R_1, R_2, R_3$ of $S$,

$$d(R_1 \cup R_2, R_3) = \min\{d(R_1, R_3), d(R_2, R_3)\}.$$

However, none of these properties can be guaranteed, unless the conditional probability structure $\{Pr_E\}_{E \subseteq \Omega}$ has the right underlying properties. For example, if the information flow between different spatial locations is asymmetrical, such as in many communications networks, then Symmetry might not be satisfied; it might take longer for information to propagate from $R$ to $R'$ than vice versa. If information can be "forgotten" or "erased" at some spatial locations in the system, then the Triangle Inequality might not be satisfied; some information propagating from $R$ to $R'$ might be forgotten before it reaches $R''$. Turning to Non-Complementarity: it is always true that $d(R_1 \cup R_2, R_3) \leq \min\{d(R_1, R_3), d(R_2, R_3)\}$. However, this inequality could be strict; i.e., we could have $d(R_1 \cup R_2, R_3) < \min\{d(R_1, R_3), d(R_2, R_3)\}$. For example, what happens in regions $R_1$ and $R_2$ at time $t_1$ could be like two pieces of a puzzle, which reveal little about what happens in region $R_3$ at time $t_2$ when considered separately, but determine it completely when put together.[54]

Note that our definition of distance between *regions* of space immediately entails a definition of distance between *points* in space: the distance between any two points $s_1$ and $s_2$ in $S$ is simply the distance between the singleton regions consisting of them, i.e., $d(s_1, s_2) = d(\{s_1\}, \{s_2\})$. Clearly, $d(s, s) = 0$ for any point $s$ in $S$. Thus, if our distance measure satisfies Symmetry and the Triangle Inequality, it determines a

---

[52] In our definition of "non-reachability", we have referred to the reference time $t_0$. However, because $\Gamma_T$ acts freely and transitively on $T$, the reference time does not matter. When region $R'$ is not reachable from region $R$ in time $t$ according to our definition, this implies that, for *any* times $t_1$ and $t_2$ with $t_2 - t_1 = t$, any event which happens in $R'$ at time $t_2$ is conditionally independent of any event which happens in $R$ at $t_1$, given $[h_{R^c \times \{t_1\}}]$.

[53] If the maximum does not exist, we can instead use the supremum, provided the order of time is *supremum-complete* (i.e., any subset of $T$ has a supremum), as it would be if $T$ were the set of real numbers. If the order of time is not supremum-complete, then the precise distance between $R$ and $R'$ may not be well-defined.

[54] Technically, this means that there exist events $E_1$, $E_2$, and $E_3$ in $\Omega$ which happen, respectively, in region $R_1$ at time $t_1$, in region $R_2$ at time $t_1$, and in region $R_3$ at time $t_2$ such that $E_1$, $E_2$, and $E_3$ are *pairwise independent*, but not *jointly independent*. This situation is common in probability theory.

*metric* on the space *S*. Furthermore, if it satisfies Non-Complementarity, this metric completely determines the distance between any two regions *R* and *R'* in *S*.[55] However, as we have pointed out, the distance measure need not generally satisfy these properties.

One notable feature of our approach is that it measures the distance between spatial locations in units of *time*. This is, of course, entirely consistent with the practice in modern physics of measuring distance in units such as *light seconds* or *light years*. However, the present approach works only if the maximum speed of information propagation in our system is finite. In classical physics, information can propagate through space at arbitrarily high speeds. Therefore, in a classical physical system, the effective "distance" between any two spatial locations collapses to zero, according to our definition. To recover a non-trivial definition of "distance" in such a system, we must impose some restriction on the sort of "information transmission" we can use. For instance, we could consider information transmission via some messenger or signal travelling at a fixed velocity. Similarly, in Maxwell's theory of electrodynamics, which is complementary to classical mechanics, electromagnetic waves propagate at a fixed and finite speed, namely the speed of light, even if classical-mechanical particles can exceed this speed. Thus, in the world of classical physics, we could define a non-trivial concept of "electromagnetic distance", even if there is no non-trivial concept of "mechanical distance". We discuss the issue of distance in quantum mechanics in Appendix C.

**4. Amorphous systems: space-time as an emergent property**

*4.1 Basic definitions*

So far, we have defined histories as functions from a set of points in either time or space-time into some state space, where histories play the role of possible worlds. Time or space-time, in turn, had an exogenously given structure. In a temporally evolving system, time was some linearly ordered set (*T*), and in a spatially extended system, space-time was explicitly decomposed into space (*S*) and time (*T*), consistent with some fixed geometry. This picture can, and for many purposes must, be generalized. Both special and general relativity theory, for example, go against the idea that there exists a fixed temporal dimension (for a classic philosophical discussion, see Putnam 1967).

A more general approach is to define a history as a function from some "indexing set", which we call a *set of loci*, into a state space. The set of loci could be a linearly ordered set of points in time, thereby accommodating our first class of systems, or a set of space-time locations with an explicit decomposition into space and time, thereby accommodating the second class. But it could also be a more general four-dimensional space-time manifold without any exogenous decomposition, or even a completely abstract indexing set.

Formally, let *I* (for "indexing set") be the *set of loci*, and let *X* denote the *state space*. A *generalized history* is a function *h* from *I* into *X*, where, for each locus *i* in *I*, *h*(*i*) is the state of the system at locus $i$. As in the case of spatially extended systems, the state $h(i)$ is best interpreted, not as a "global" state, in which the system is at some specific point in time (indeed, there is no exogenous notion of time), but as a "local" state, in which the system is at a specific locus. We write $\Omega$ to denote the set of all generalized histories deemed *possible*, which can again be viewed as nomologically possible worlds, and subsets of $\Omega$ are called *events*.[56]

---

[55] To be precise, $d(R,R') = \min\{d(s,s'): s \in R \text{ and } s' \in R'\}$. Strictly speaking, this only works if *R* and *R'* are *finite* sets of points. For *infinite* sets, we would need a slightly stronger version of non-complementarity, which says that $d(R, R') = \inf\{d(s, s'): s \in R \text{ and } s' \in R'\}$ (and this infimum exists).

To complete our present definitions, we must, once more, introduce a *conditional probability structure* on $\Omega$. As should be clear by now, this is a family of conditional probability functions $\{Pr_E\}_{E \subseteq \Omega}$, consisting of one $Pr_E$ for each event $E$ in $\Omega$. An *amorphous system* is the pair consisting of the set $\Omega$ of possible generalized histories and the conditional probability structure $\{Pr_E\}_{E \subseteq \Omega}$.

How much of the framework that we have developed so far can be extended to the setting of amorphous systems? We might ask, for instance, whether an abstract indexing set, despite not being endowed with any exogenous structure, can attain some spatial and/or temporal structure as an emergent property, for instance as a byproduct of the correlations encoded in $\{Pr_E\}_{E \subseteq \Omega}$. We might also ask whether, and to what extent, the geometry of the set of loci is unique, or alternatively whether there might be multiple equally "correct" geometries.

*4.2 Adjacency structure and the Markov property*

Just as in Section 3.8, the topology of the set $I$ of loci can be represented by an *adjacency structure*: a binary relation $\rightarrow$ defined between subsets of $I$. For example, suppose $I$ is a set of times, as in Section 2, i.e., $I = T$. For any subsets $R$ and $R'$ of $I$, define $R \rightarrow R'$ if there does not exist any time $t$ such that $r < t < r'$ for all $r$ in $R$ and all $r'$ in $R'$. For another example, let $I$ be the four-dimensional space-time manifold of a general relativistic system. Then, for any subsets $R$ and $R'$ of $I$, we might define $R \rightarrow R'$ if there is a locus $i$ in $I$ such that any open neighbourhood around $i$ intersects $R'$.

In Section 2.10, we related the order structure of the set $T$ of times to the conditional probability structure $\{Pr_E\}_{E \subseteq \Omega}$ by means of a temporal Markov property. Likewise, in Section 3.8, we related the adjacency structure of the set $S$ of spatial locations to the conditional probability structure $\{Pr_E\}_{E \subseteq \Omega}$ by means of a spatial Markov property. We now discuss a similar idea in relation to a general set of loci. This, in turn, will allow us to view the adjacency structure among loci – and thereby its topology – as an "emergent property": something that emerges from the correlations encoded in $\{Pr_E\}_{E \subseteq \Omega}$.

Let $R$ be a subset of $I$ (i.e., a collection of loci). As before, for any generalized history $h$ in $\Omega$, we define $h_R$ to be the restriction of that history to the set $R$. We then define the event $[h_R]$ to be the set of all histories $h'$ in $\Omega$ such that $h_R = h'_R$. For any event $E$ (i.e., a subset of $\Omega$), we say that *E happens inside R* if, for all histories $h$ and $h'$ such that $h_R = h'_R$, the history $h$ is an element of $E$ if and only if $h'$ is an element of $E$. Thus, whether or not a particular history is an element of $E$ is completely determined by the restriction of that history to $R$.

---

[56] Note that, in the literature on general relativity theory, the word "event" is used to refer to the objects we call "loci". Our use of the word "event" is consistent with its use in probability theory.

As in Section 3.8, we define a *tripartition* of the set *I* of loci to be a triple (*R*, *R'*, *R''*), where *R*, *R'*, and *R''* are three disjoint subsets of *I* which together cover *I* (i.e., $R \cup R' \cup R'' = I$), such that it is *not* the case that $R \to R''$ or $R'' \to R$. Again, this means that the set *R'* somehow "separates" *R* from *R''*.

For example, let *T* be a set of times, with the adjacency structure introduced at the start of this section. Fix two times $t_0$ and $t_1$, with $t_0 \leq t_1$. Let *R* be the set of all times *strictly before* $t_0$, let *R'* be the set of all times *between* $t_0$ and $t_1$ (including $t_0$ and $t_1$), and let *R''* be the set of all times *strictly after* $t_1$. Then (*R*,*R'*,*R''*) is a tripartition of *T*.

For another example, let *I* be the four-dimensional Minkowski space-time of special relativity, with the "open neighbourhood" adjacency structure introduced above. Let λ be a linear time-like trajectory through *I*, for instance the trajectory of an "observer" traveling through space-time at a constant velocity, and let *p* be a point on this trajectory. In special relativity theory, there is a unique three-dimensional *simultaneity hyperplane R'* passing through *p*, such that all events that happen inside *R'* seem to occur simultaneously from the perspective of the λ-observer at *p*. Let *R* be the set of all points in *I* which have some part of *R'* in their *future* light-cone, and let *R''* be the set of all points in *I* which have some part of *R'* in their *past* light-cone. Then (*R*, *R'*, *R''*) is a tripartition of *I*.[57] More generally, let *R* and *R''* be any disjoint open subsets of *I*,[58] and let *R'* be the complement of the union $R \cup R''$. Then (*R*, *R'*, *R''*) is a tripartition of *I*.

We say that the adjacency structure → satisfies the *amorphous Markov property* with respect to the conditional probability structure $\{Pr_E\}_{E \subseteq \Omega}$ if, for any tripartition (*R*, *R'*, *R''*) and any generalized history *h* in Ω, any event which happens inside *R* is conditionally independent from any event which happens inside *R''*, given $[h_{R'}]$. Again, this means, roughly, that there is no way for information to propagate from *R* into *R''*, or vice versa, without first passing through *R'*. For example, suppose *I* is four-dimensional Minkowski space-time, and (*R*, *R'*, *R''*) is the tripartition described above. In this case, the simultaneity hyperplane *R'* plays the role of the "present", which isolates the "past" *R* from the "future" *R''*. If we have complete information about the history inside *R'* (i.e. we know $[h_{R'}]$), then we have complete information about the "present state" of the world; thus, we can predict its future evolution (in *R''*) without needing to know anything about its past history (in *R*).

In Section 2.10, we argued that the temporal Markov property was the key property of time; a "correct" ordering of the set *T* was any ordering that satisfied this property. Likewise, in Section 3.8, we argued that the spatial Markov property was the key property of space; a "correct" adjacency structure on the set *S* was any adjacency structure that satisfied this property. Now we make a parallel claim for amorphous systems: a "correct" adjacency structure on *I* is one that satisfies the amorphous Markov property. This Markov property subsumes both the temporal Markov property of Section 2 and the spatial Markov property of Section 3.

---

[57] In a model of general relativity, a similar construction works if *R'* is a Cauchy surface in the four-dimensional space-time manifold.

[58] A subset *R* of *I* is *open* if, for any *s* in *R*, there is some $r > 0$ such that the ball of radius *r* around *s* is contained in *R*.

This has an important consequence. In our framework, the topology of *I*, in the form of the adjacency structure, does not need to be imposed exogenously. Instead, this topology can emerge *endogenously* from the conditional probability structure $\{Pr_E\}_{E \subseteq \Omega}$. We say that an adjacency structure $\rightarrow$ between subsets of *I* is $\{Pr_E\}_{E \subseteq \Omega}$-*admissible* if it satisfies the amorphous Markov property with respect to $\{Pr_E\}_{E \subseteq \Omega}$. If we think of *I* as a sort of generalized space-time, this means that the topology of space-time is an emergent property of the amorphous system.[59]

*4.3 Time and predictability*

Both temporally evolving systems and spatially extended systems come with a set *T* which plays the role of "time". What plays the role of time in an amorphous system? The adjacency structure described in the previous section tells us whether two subsets of the index set *I* are in "informational contact" or are "informationally separated" from one another, but it does not tell us which subset comes "before" and which comes "after", or even whether this question makes sense. We now explain how time itself can be an emergent property of an amorphous system.

Let $\rightarrow$ be an adjacency structure on the index set *I*. Let *T* be a linearly ordered set. A *possible time structure* on *I* is a function $\tau$ from *I* into *T* such that, for any *t* in *T*, if (i) *R* is the set of all points *i* in *I* such that $\tau(i) < t$, (ii) *R'* is the set of all points *i* in *I* such that $\tau(i) = t$, and (iii) *R''* is the set of all points *i* in *I* such that $\tau(i) > t$, then (*R*, *R'*, *R''*) is a tripartition of *I*.

For example, let *I* be four-dimensional Minkowski space-time as described in Section 4.2, and let $\lambda$ be a linear time-like trajectory through *I*. Fix some point $p_0$ on the trajectory $\lambda$. Let *T* be the set of real numbers. Then, for every *t* in *T*, there is a unique point $p_t$ along the trajectory $\lambda$ which appears to be *t* seconds in the future of $p_0$ (or in the past, if $t < 0$), with respect to the subjective time (i.e., *proper time*) experienced by an observer traveling along the trajectory $\lambda$. Let $R_t$ be the simultaneity hyperplane passing through $p_t$. If we define $\tau(i) = t$ for all points *i* in $R_t$, then $\tau$ is a possible time structure on *I*.

As this example illustrates, an amorphous system may admit *many* possible time structures. In special relativity, there is a distinct time-structure for every inertial reference frame. All of these time structures are equally "correct". Indeed, this is one of the key insights of special relativity theory. However, unless we impose further constraints, a system may also admit many "absurd" time structures. For example, suppose *I* is four-dimensional Newtonian space-time (i.e., $I = \mathbf{R}^3 \times \mathbf{R}$), with the "open neighbourhood" adjacency structure described in Section 4.2. For all points $(s_1, s_2, s_3, t)$ in $I$, define $\tau(s_1, s_2, s_3, t) = s_3$. Then $\tau$ is a possible time structure on $I$. But if the "true" time coordinate is $t$, not $s_3$, it seems that this time structure is not correct. So, what property of the system determines which time structures are the correct ones?

---

[59] We are not the first to suggest that the geometry and/or topology of space-time could be an emergent property of more fundamental causal structures. Brown and Pooley (2001, 2006) have argued forcefully that the geometry – or rather, the *apparent* geometry – of relativistic space-time should be seen as a consequence of the symmetries (i.e., Lorentz covariance) of the dynamical laws governing matter and electromagnetism. In their words (2006, Section 5): "space-time's Minkowskian structure cannot be taken to explain the Lorentz covariance of the dynamical laws. From our perspective, of course, the direction of explanation goes the other way around. It is the Lorentz covariance of the laws that underwrites the fact that the geometry of space-time is Minkowskian." See also Brown (2005). However, Brown and Pooley's approach is very different from the approach we take here. The idea of emergent space-time geometry has also recently appeared in the literature on high-energy physics and quantum cosmology. See, e.g., Konopka et al. (2008) and Hamma et al. (2010).

Clearly, a "correct" time structure should satisfy something like the temporal Markov property from Section 2. However, if the adjacency structure $\rightarrow$ satisfies the amorphous Markov property with respect to the conditional probability structure $\{Pr_E\}_{E \subseteq \Omega}$, then it is easy to see that *any* possible time structure will satisfy the temporal Markov property.[60] So, the Markov property alone is not enough to pick out the "correct" time structures.

Arguably, what picks out the correct time structures is *predictability*. To understand this, suppose we took a classical mechanical system with Newtonian space-time $I = \mathbf{R}^3 \times \mathbf{R}$, and applied the "absurd" time structure $\tau(s_1, s_2, s_3, t) = s_3$, as defined above. How would the system appear with respect to this time structure? It would appear very strange and unpredictable. Particles would randomly pop in and out of existence. Energy and momentum would not be conserved from one moment to the next. Events would seem to unfold over time without any rhyme or reason. This total lack of predictability would be an indication that we had picked the wrong time structure for the system.

On the other hand, if we had picked the "correct" time structure, namely $\tau(s_1, s_2, s_3, t) = t$, then the system would appear completely deterministic; its state at one "moment" in time, as defined by $\tau$, would completely determine its "past" and "future" behaviour, as defined by $\tau$. This total predictability is an indication that this is the correct time structure for the system.

In this example, there was a particularly stark contrast between an "incorrect" time structure, which renders the system totally unpredictable, and a "correct" one, which renders it totally predictable. This is because classical mechanical systems are deterministic. In an indeterministic system, there will not generally be such a stark contrast. Nevertheless, some time structures will render the system more predictable than others, and among these, we claim, the ones that render the system *most predictable* are the correct time structures for that system.

To make this idea more precise, we need a way to measure the "predictability" of a system under a given time structure. One way to do this is to use the information-theoretic concept of *entropy*.[61] For any subset $R$ of $I$, let $\Omega_R$ be the set of all $R$-restricted histories $h_R$ obtained from any $h$ in $\Omega$. For simplicity, let us assume that the underlying state space $X$ is finite. If $R'$ is some other finite subset of $I$, then $\Omega_{R'}$ is also finite.[62] Suppose we know $h_R$, and we want to predict $h_{R'}$. For any $h_R$ in $\Omega_R$, there is a quantity called the *conditional entropy of R' given* $h_R$, denoted by $\eta(R',h_R)$, which measures how "unpredictable" the restricted history $h_{R'}$ is, given the restricted history $h_R$.[63] For example, if $h_{R'}$ is entirely determined by $h_R$, then $\eta(R',h_R) = 0$. At the other extreme, if $h_{R'}$ is effectively as unpredictable as a collection of independent coin-tosses, even after conditioning on $h_R$, then $\eta(R',h_R) = 1$. Intermediate levels of entropy represent intermediate degrees of unpredictability.

---

[60] To be somewhat more precise: given a possible time structure on $I$, we can represent the amorphous system as a temporally evolving system, and this temporally evolving system will satisfy the temporal Markov property. This construction is straightforward, but to avoid getting bogged down in technicalities, we set aside the details here.

[61] This is *not* the same as thermodynamic entropy, although it is loosely related. Thus, the discussion that follows should *not* be interpreted thermodynamically.

[62] To be precise, if $|X|$ is the cardinality of $X$, and $|R'|$ is the cardinality of $R'$, then the cardinality of $\Omega_{R'}$ is at most $|X|^{|R'|}$.

Now, let $\tau$ be a time structure, mapping $I$ into $T$. Let $t$ be some time in $T$; let $R$ be the set of all points $i$ in $I$ such that $\tau(i) = t$; and let $R^C$ be the set of all points $i$ in $I$ such that $\tau(i) \neq t$. We define $\eta(\tau, t)$, the *unpredictability* of the system under $\tau$ at $t$, to be the maximum value of $\eta(R', h_R)$, where $h_R$ can be any element of $\Omega_R$ and $R'$ is allowed to be any finite subset of $R^C$.[64] If $\eta(\tau, t)=0$, then this means roughly that any generalized history $h$ in $\Omega$ is almost entirely predictable, based on its restriction $h_R$.[65] If $\eta(\tau,t) > 0$, then histories in $\Omega$ are not, in general, fully predictable from their restrictions to $R$. The larger $\eta(\tau, t)$ is, the less predictable these histories are. We then define $\eta(\tau)$, the *unpredictability* of the system under the time structure $\tau$, to be the maximum value of $\eta(\tau, t)$ over all times $t$ in $T$.[66]

For example, suppose $I$ is the four-dimensional Newtonian space-time of a classical mechanical system (i.e., $I = \mathbf{R}^3 \times \mathbf{R}$), and $\tau$ is the "correct" time structure for this system, namely $\tau(s_1,s_2,s_3,t) = t$. Then $\eta(\tau) = 0$, because classical mechanics is entirely deterministic. However, if $\tau$ was an "incorrect" time structure, such as $\tau(s_1,s_2,s_3,t)=s_3$, then we would have $\eta(\tau) > 0$, because the ascription of this incorrect time structure would render the system unpredictable, as we have explained.

We now come to the key point of this section. A *correct* time structure for an amorphous system is any time structure that *minimizes* unpredictability and thereby maximizes regularity. Note that this definition allows that there may be many correct time structures, as in the case in special or general relativity, all of which render the system equally predictable. This has an important consequence. The correct time structure does not need to be imposed exogenously. Instead, the correct time structure (or structures) could emerge *endogenously* from the conditional probability structure $\{Pr_E\}_{E \subseteq \Omega}$. In other words, the structure of time itself could itself be an emergent property of the amorphous system. Using a more metaphysical language, it might be that space and time are grounded in the dynamics of the system, rather than the other way round.

---

[63] Formally, $\eta(R', h_R)$ is the sum, over all possible $R'$-restrictions $h_{R'}$ in $\Omega_{R'}$, of
$$-Pr([h_{R'}] \mid [h_R]) \log_2[Pr([h_{R'}] \mid [h_R])] / |R'| \log_2(|X|).$$
However, the precise formula is not important for this discussion.

[64] To be more precise, it is the *supremum* of this set. The maximum is not always well-defined.

[65] Even if $\eta(\tau, t)=0$, there may be some "residual" unpredictability, in the sense that $\Omega$ may contain more than one extension of $h_R$ to all of $I$. However, the conditional probability structure $\{Pr_E\}_{E \subseteq \Omega}$ concentrates all probability on *one* of these possible extensions; the rest of the extensions get probability zero.

[66] Again, strictly speaking, we require the supremum. The supremum of $\eta(\tau, t)$ across time $t$ is not the only conceivable measures of unpredictability of the system under time structure $\tau$. We could also take the average or some other aggregate measure. For example, suppose that $I$ is an $N$-dimensional integer lattice (formally, $I = \mathbf{Z}^N$). Then we could measure the unpredictability of the system under different time structures using the theory of *entropy geometry* and *expansive subdynamics* first developed for multidimensional cellular automata by Milnor (1988) and later extended to arbitrary multidimensional symbolic dynamical systems by Boyle and Lind (1997). See the section on "Entropy" in Pivato (2009) for a summary of this theory.

*4.4 Which features of a system are real?*

A final philosophical question on which we wish to comment briefly is the following. Suppose we have described a given system using our formal framework. Should we treat all features of that system as "real", or should we treat some features as mere artefacts of our formal description?

The debates between *relationalist* and *substantivalist* views about space and time, and between *structuralist* and *full-blown realist* views in science more generally, can be seen as attempts to answer this question.[67] Let us begin with a *relationalist* or *structuralist* view, which may be about space and time in particular or about the properties of a system more generally. On such a view (of which there can be several variants), only some "relational" or "structural" properties of a system should be viewed as real, while "intrinsic", "non-structural" properties should not. It does not matter, for example, what the nature of the system's spatiotemporal loci in the set $I$ is, nor what the nature of the system's possible states in the set $X$ is. All that matters is how these loci and/or states are related to one another and what dynamics they display. Two formally distinct systems, with formally distinct indexing sets $I$ and $I'$ and/or formally distinct state spaces $X$ and $X'$, will count as the same system if their nomologically possible histories and probabilistic properties are structurally indistinguishable.

By contrast, on a *substantivalist* or *full-blown realist* view, which may also be about space and time in particular or about the properties of a system more generally, even intrinsic, non-structural properties of a system can be real, over and above the system's relational or structural properties. So, the system's spatiotemporal indexing set $I$ and its state space $X$ may be significant in ways that go beyond the structures and relations in which they stand. (Again, there can be several variants of such a view.) An example of a non-structural property is the exact indexing of time. One can imagine two structurally identical temporally evolving systems, indexed by $T = \{0,1,2,3,....\}$ and $T' = \{1,2,3,4,....\}$ respectively. The only difference is that in one system history "starts at time zero", whereas in the other it "starts at time 1". For a relationist or structuralist, these are "the same" system. But a substantivalist or full-blown realist might insist that there is a genuine difference between them.

The debates between these different views occur in several places in philosophy and take a variety of forms, so we cannot do justice to them here. We wish to note, however, that our formal framework can be used to express some salient positions within those debates. Specifically, different answers to the question of which features of a system are real can be expressed in terms of different criteria for individuating systems. If we begin with a very large class of systems that are formally described in our framework, there are a number of ways in which one might partition this class of systems into equivalence classes that are each taken to represent the same system. Different such partitions then correspond to different answers to the question of which features of a system are real, rather than mere artefacts of our formal description. In particular, only those features that are present among *all* members of any given equivalence class count as real. Features on which there can be differences even *within* the same equivalence class count as artefacts of our formal description.

---

[67] On a broadly "structuralist" or "relationalist" approach to metaphysics, see, e.g., Ladyman and Ross (2009). On "absolute" versus "relational" accounts of space and time, see, e.g., Earman (1989). On "substantivalism" and its discontents, see, e.g., Nerlich (2003).

A relationalist or structuralist view would entail that any two systems that do not differ in any relational or structural properties count as the same and thereby fall into the same equivalence class. A substantivalist or full-blown realist view, by contrast, would entail that two such systems could still count as different; thus, the equivalence classes would be more fine-grained according to such a view, and might even be singleton (in which case *all* features of any given system would count as real).

Here is one way of formalizing this idea. Consider two amorphous systems, given by the pairs $(\Omega, \{Pr_E\}_{E \subseteq \Omega})$ and $(\Omega', \{Pr'_E\}_{E \subseteq \Omega'})$, where the histories in $\Omega$ are functions from the set $I$ of loci into the state space $X$, and the histories in $\Omega'$ are functions from the set $I'$ of loci into the state space $X'$. Let $\mathcal{H}$ and $\mathcal{H}'$ denote the sets of logically possible functions from $I$ into $X$ and from $I'$ into $X'$, respectively.

Suppose there is a bijection $\theta$ from $I$ into $I'$, and also a bijection $\xi$ from $X$ into $X'$ (recall that a *bijection* is a one-to-one, onto function). Using $\theta$ and $\xi$, we can then define a bijection $\sigma$ from $\mathcal{H}$ into $\mathcal{H}'$ which maps each history $h$ in $\mathcal{H}$ to the history $h'$ in $\mathcal{H}'$ defined as follows: for each $i'$ in $I'$,

$h'(i') = \xi[h(i)]$, where $i = \theta^{-1}(i')$ (with $\theta^{-1}$ defined as the inverse of $\theta$).

The bijection $\sigma$ is an *isomorphism* between the two systems if

- $\sigma(\Omega) = \Omega'$; and
- for any events $E'$ and $D'$ in $\Omega'$, if $E$ and $D$ are the inverse images of $E'$ and $D'$ under $\sigma$, then $Pr'_E(D') = Pr_E(D)$.

We call two systems *isomorphic* if there exists an isomorphism between them. Isomorphic systems display the same dynamics, and they are relationally or structurally indistinguishable.[68] Moreover, any topology of space and time that is admissible for one such system can be mapped, in a structure-preserving way, onto a topology that is admissible for another.

Thus, on a relationalist or structuralist view, any two isomorphic systems should be considered the same. On a substantivalist or full-blown realist view, they may still differ. A view of the first kind would therefore take systems to be unique only up to isomorphism, so that our initial large class of systems would be partitioned into equivalence classes of isomorphic systems. A view of the second kind would opt for a more fine-grained partition, acknowledging that even isomorphic systems may be distinct in reality.

The properties of systems on which we have focused in this paper are mainly structural and are preserved by all isomorphisms. This includes, for instance, the symmetries and ergodicity properties of a system, the distinction between laws and

---

[68] In fact, any bijective symmetry of a system constitutes an isomorphism from a system into itself.

"brute facts", and the topology (or topologies) and geometry (or geometries) of space and time that are compatible with the correlation structure (in the sense that they satisfy the relevant Markov conditions). Thus, even a relationalist or structuralist would accept that all of these properties are "real" features of the system, and not mere artefacts.

**5. Concluding discussion**

We have introduced a framework for describing three general classes of systems and shown how this framework can be used to address a number of philosophical questions. We began with the class of temporally evolving systems, of which classical dynamical systems are a special case, and then moved on to the class of spatially extended systems and the class of amorphous systems. As noted, the framework can accommodate systems as diverse as the solar system, quantum-mechanical systems, special and general relativistic systems, and the earth's climate system.

We have discussed questions such as: how can we define nomological possibility, necessity, determinism, and indeterminism? How can we distinguish laws from brute facts? How are laws related to symmetries? What regularities must a system display to permit global generalizations from local observations? How can we formulate principles of parsimony such as Occam's Razor, and how can we justify their use? What is the role of space and time in a system? And what is at stake in the debate between relationalist and substantivalist views about space and time, and between structuralist and full-blown realist views about systems more generally?

While our framework and what it says about these questions should already be of sufficient interest to make the framework worth studying, the greatest payoff lies arguably in the variety of applications to which the framework lends itself. Developing these is beyond the scope of this paper, but we conclude by mentioning a few.

*5.1 Higher-level versus lower-level properties*

Our framework can be used to explore the relationship between lower-level ("micro") and higher-level ("macro") properties of a system. By partitioning the system's state space $X$ into suitable equivalence classes, we can capture the idea that "higher-level" or "macro" states are more coarse-grained than "lower-level" or "micro" states, so that each "macro" state can be realized by multiple "micro" states: the phenomenon of *multiple realizability*. Consider, for example, all the different possible micro-level trajectories of a tossed coin that each correspond to the macro-property of "landing heads". Or consider all the different possible micro-states of individual water molecules that each correspond to a macro-state such as "frozen", "liquid", or "gaseous".

Suppose $X$ is the original state space, and $\mathbb{X}$ is the relevant set of equivalence classes, which we interpret as the higher-level state space. We can then write σ to denote the function that maps each lower-level state $x$ in $X$ to the corresponding higher-level state $\mathbb{x}$ in $\mathbb{X}$. (Note the "outlined" font for higher-level objects.) This function can be interpreted as the *supervenience relation* connecting the two levels. We can then use σ to specify the resulting higher-level histories.[69] For each lower-level history $h$ in the original set $\Omega$, the corresponding higher-level history $\hbar$ is the function from $T$ into $\mathbb{X}$, where, for each $t$ in $T$, $\hbar(t) = \sigma(h(t))$. (If we are dealing with a spatially extended or amorphous system instead of a temporally evolving one, we must replace $T$ in this definition with $S \times T$ or $I$.) The set of higher-level histories is therefore $\mathcal{Q} = \sigma(\Omega)$. Similarly, we can use σ to arrive at a conditional probability structure defined over higher-level events, formally written $\{\mathcal{P}r_E\}_{E \subseteq \mathcal{Q}}$; see Appendix A for details. The pair $(\mathcal{Q}, \{\mathcal{P}r_E\}_{E \subseteq \mathcal{Q}})$ can be viewed as our system, *re-described at a higher level*. In the terminology of Appendix A, the higher-level system is a *factor system* of the original, lower-level system.

This construction allows us to study the dynamics of the higher-level system and to compare its properties with those of the lower-level system. Interestingly, the higher-level dynamics may be different from the underlying lower-level dynamics. For example, features such as determinism or indeterminism are not generally preserved under coarse-graining: the lower-level system may be deterministic, while the higher-level system is not (or vice versa). Thus indeterminism could be an emergent property (see, e.g., Butterfield 2012 and List 2014; for a related discussion, see also Werndl 2009b).

In a similar vein, we may study the level-specificity of other properties. For instance, elsewhere we have used this approach to argue that non-trivial objective chance could be an emergent phenomenon, which is entirely consistent with lower-level determinism (List and Pivato 2015).[70]

*5.2 Laws and regularities in the special sciences*

There is much debate on whether there are laws in the special sciences, as distinct from fundamental physics. The existence of laws is particularly contested in fields such as biology, ecology, geology, psychology, and the social sciences. (Chemistry, by contrast, is often viewed as a close relative of physics and thereby similar enough to it in its lawfulness.) Examples of special-science regularities that are sometimes described as laws include (i) Kleiber's law in biology, according to which an organism's metabolic rate is proportional to the ¾[th] power of its body mass; (ii) the laws of supply and demand in economics, according to which (except for Giffen goods) the demand for a good is a decreasing function of its price, and the supply is an increasing function of price; and (iii) Duverger's law in political science, according to which, under a first-past-the-post electoral system, the effective number of parties in the legislature will be lower than under a proportional-representation system, *ceteris paribus*. The key question is whether any of these regularities are sufficiently robust to qualify as laws.

One common view is that, as we move further away from fundamental physics, there are fewer and fewer regularities that live up to genuine "law-hood". Kim (2010, ch. 14), for instance, argues that there are no "strict" laws in the special sciences. Among the reasons he gives for this conclusion are (i) the multiple realizability of special-science properties, which, he claims, undermines their "inductive projectibility", and (ii) the alleged metaphysical anomalism of the mental realm, which, he suggests, undermines the existence of laws in psychology and the social sciences.

---

[69] This construction, under the present notational conventions, was introduced in List (2014) and List and Pivato (2015).
[70] For earlier work defending higher-level chance, sometimes using a strategy that is similar in spirit to ours (though not fully equivalent), see, e.g., Loewer (2001), Frigg and Hoefer (2010), Glynn (2010), Strevens (2011), and Hemmo and Shenker (2012).

Other scholars defend the existence of laws in the special sciences. For example, focusing on the social sciences, Kincaid (1990) argues that several widely cited arguments against laws fail. He thinks that the most serious challenge to laws in the social sciences comes from the excessive *ceteris paribus* qualifications that all such laws require, but argues that the procedures we routinely employ to deal with such qualifications in the natural sciences carry over to the social sciences.

The framework we have presented might be used to make some progress in this debate. Using our framework, we can in principle describe the special-science systems in question and identify the properties these systems would have to display in order to secure the existence of laws. As we have seen, what laws there are in a given system depends on the system's symmetries and the properties they preserve. Another question is whether we are prepared to recognize weaker kinds of laws corresponding to *partial* or *local* symmetries, as defined in Appendix B. Finally, our framework shows that whether we can make global generalizations from local observations, and thereby come to know the relevant laws, depends on whether the given special-science systems are ergodic.[71] Although it is undoubtedly difficult to settle all of these questions, our framework can make them explicit, thereby rendering the debate more tractable.

*5.3 Intentional systems*

While we have mainly discussed physical systems, there is no barrier, in principle, to using our framework also for describing systems involving intentional agents. Indeed, an earlier version of the present formalism has proved useful for the analysis of free will (List 2014 and List and Rabinowicz 2014). We can think of an agent, together with the relevant environment, as a temporally evolving system. This system can be described at different levels: at a physical level, at which we would *not* take an "intentional stance" towards the system, and at an agential level, at which we *would* take such a stance (on the notion of an "intentional stance", see Dennett 1987). Physical-level descriptions would capture the states of the agent's brain and body, while agential-level descriptions would capture the agent's higher-level mental or psychological states, thereby focusing on the agent's beliefs, desires, and intentions, rather than the underlying neuronal or bodily states.

The framework then allows us to explain, for instance, how agential-level indeterminism and an agent's *possibility of doing otherwise* can co-exist with physical-level determinism (as argued in List 2014). The framework might also shed some light on how other psychological properties can emerge from the underlying physical dynamics of the system. In particular, as a factor system of the original physical system, the agential system may exhibit additional symmetries not present at the physical level. This may, in turn, be used to explain why some higher-level regularities in an intentional system (e.g., regularities involving beliefs, desires, intentions, and norms) may qualify as "real patterns" (Dennett 1991) and not merely as illusions due to our ignorance of the physical-level details.

---

[71] The systems that are studied in the special sciences often arise as higher-level descriptions of systems from the natural sciences, as discussed in Section 5.1 above. In this case, the special-science system will have at least as many symmetries and at least as much ergodicity as the underlying natural-science system, as explained in Appendix A.

Needless to say, all of these applications are challenging and raise controversial philosophical issues. We hope, however, that our framework will be a clarifying contribution to formal metaphysics and the philosophy of science and will inspire further work.

**Appendix A: Factor systems**

One possible objection to our framework is that it is both unrealistic and unwieldy. It is unrealistic because the actual universe is not sufficiently regular (e.g., it might lack a large enough monoid of symmetries or ergodicity). It is unwieldy because the universe as a whole is far too complex a system for us to analyze within this framework anyway.

However, there is no need to insist on applying our framework to the universe as a whole. Instead, we can apply it to a "factor" system, as we now explain. Consider a temporally evolving system, consisting of a set $\Omega$ of nomologically possible histories (each of which is a function from a set $T$ of times into a set $X$ of possible states), along with a conditional probability structure $\{Pr_E\}_{E \subseteq \Omega}$. Let $X'$ be another set, and let $\phi$ be a function from $X$ into $X'$. For every history $h$ in $\Omega$, let $\phi(h)$ be the function $h'$ from $T$ into $X'$ defined by $h'(t) = \phi[h(t)]$ for all $t$ in $T$. Let $\Omega' = \{\phi(h): h \in \Omega\}$. For any subsets $D'$ and $E'$ of $\Omega'$, let $D$ and $E$ be their inverse images under $\phi$, where these are subsets of $\Omega$, and define $Pr'_{E'}(D') = Pr_E(D)$. Then $\{Pr'_{E'}\}_{E' \subseteq \Omega'}$ is a conditional probability structure on $\Omega'$. The system specified by $\Omega'$ and $\{Pr'_{E'}\}_{E' \subseteq \Omega'}$ is called a *factor system* of the original system specified by $\Omega$ and $\{Pr_E\}_{E \subseteq \Omega}$. The function $\phi$ from $\Omega$ into $\Omega'$ is called a *factor map*. We can define *factors* of spatially extended systems and amorphous systems in an exactly analogous manner; we leave the details to the reader.

For a concrete example, suppose that ($\Omega$, $\{Pr_E\}_{E\subseteq\Omega}$) is a classical-physics description of the entire solar system at an atomic level of detail. So $X$ is an extremely high-dimensional space, which must specify the position and momentum of every atom in the entire solar system, along with all of their gravitational and electromagnetic interactions.[72] Meanwhile, let ($\Omega'$, $\{Pr'_{E'}\}_{E'\subseteq\Omega'}$) be the very simple celestial-mechanical system consisting only of the Earth, Moon, and Sun, described as gravitationally interacting point masses. So $X' = \mathbf{R}^{18}$, because we must specify the three-dimensional position and momentum vectors of each of the three objects in the system, and 3×6 = 18. Let $\phi$ be the function from $X$ into $X'$ which translates each highly detailed atomic-level description of the solar system into the crude 18-dimensional celestial mechanical description. Then $\phi$ is a factor mapping from ($\Omega$, $\{Pr_E\}_{E\subseteq\Omega}$) into ($\Omega'$, $\{Pr'_{E'}\}_{E'\subseteq\Omega'}$), and thus ($\Omega'$, $\{Pr'_{E'}\}_{E'\subseteq\Omega'}$) is a factor of ($\Omega$, $\{Pr_E\}_{E\subseteq\Omega}$).

As this example illustrates, a factor system can be seen as a sort of "abstraction" or "simplification" of the original system, obtained by discarding some properties. Now, suppose $\psi$ is a function from $T$ into itself (e.g., a time shift) which is a temporal symmetry of the original system ($\Omega$, $\{Pr_E\}_{E\subseteq\Omega}$). Then it is easy to verify that $\psi$ will also be a temporal symmetry of the factor system ($\Omega'$, $\{Pr'_{E'}\}_{E'\subseteq\Omega'}$). Thus, the temporal symmetry monoid of the factor ($\Omega'$, $\{Pr'_{E'}\}_{E'\subseteq\Omega'}$) is *at least as large* as the temporal symmetry monoid of the original system ($\Omega$, $\{Pr_E\}_{E\subseteq\Omega}$). In a spatially extended system, the exact same statement applies to spatiotemporal symmetries. Furthermore, if $\Psi$ is an amenable group of temporal (or spatiotemporal) symmetries, and ($\Omega$, $\{Pr_E\}_{E\subseteq\Omega}$) is ergodic relative to $\Psi$, then ($\Omega'$, $\{Pr'_{E'}\}_{E'\subseteq\Omega'}$) will *also* be ergodic relative to $\Psi$. In other words, ($\Omega'$, $\{Pr'_{E'}\}_{E'\subseteq\Omega'}$) is *at least as ergodic* as ($\Omega$, $\{Pr_E\}_{E\subseteq\Omega}$).

This means that, even if the original system ($\Omega$, $\{Pr_E\}_{E\subseteq\Omega}$) lacks certain symmetries or ergodicity properties, the factor system ($\Omega'$, $\{Pr'_{E'}\}_{E'\subseteq\Omega'}$) may well possess these properties. Furthermore, even if the original system ($\Omega$, $\{Pr_E\}_{E\subseteq\Omega}$) is too complicated to analyze using the formal tools we have described, the system ($\Omega'$, $\{Pr'_{E'}\}_{E'\subseteq\Omega'}$) may well be simple enough. To illustrate this, consider our example of the solar system. The original system ($\Omega$, $\{Pr_E\}_{E\subseteq\Omega}$) describes the entire solar system at an atomic level of detail; whether or not the system possesses the desired symmetries or ergodicity properties, it is certainly too complex to analyze. In contrast, the abstract Earth-Moon-Sun system ($\Omega'$, $\{Pr'_{E'}\}_{E'\subseteq\Omega'}$) is very simple. In fact, it is an example of a *quasiperiodic* dynamical system: it can be described as two independently rotating "wheels", one describing the orbit of the Moon around the Earth, and the other describing the orbit of the Earth around the Sun. This is a prototypical example of an ergodic dynamical system.

**Appendix B: Partial symmetries and local symmetries**

An important assumption of this paper has been that there is a fairly large monoid $\Gamma$ of symmetries acting on the set $\Omega$ of nomologically possible histories. We have argued that a modal or probabilistic feature of the system is a physical "law" rather than a "brute fact" if it is invariant under all of these symmetries. But this argument runs into a problem: many systems studied in the sciences lack sufficient symmetries to account for all of their "law-like" features.

---

[72] For simplicity, we eschew a quantum-mechanical description in this example.

For example, suppose that space is represented by the set of all integers, while time is represented by the set of positive integers, i.e., $S = \{...,-1,0,1,2,...\}$ and $T = \{1,2,3,...\}$, and consider the simple random-walk system described in Section 2.10. Nomologically speaking, the token could begin at any spatial location at time one. But suppose the conditional probability structure $\{Pr_E\}_{E \subseteq \Omega}$ is such that, with probability one, the token begins at spatial location zero at time one.[73]

In that case, the probability distribution of its location at time $t$ is a $(t–1, ½)$-*binomial distribution*.[74] Evidently, this distribution is not invariant under spatial translations, since it is centred around zero. Furthermore, it changes over time. Thus, spatiotemporal translations are not symmetries of this system. But this contradicts our intuition that the motion of the token is highly "law-like": it can be described by a simple rule which is the same everywhere in space and time.

To solve this problem, we now introduce the notion of "partial" symmetries; and we adopt the framework of spatially extended systems. Recall that $\mathcal{H}$ is the set all logically possible spatially extended histories. A *partial symmetry monoid* of a spatially extended system is a collection Γ of transformations of $\mathcal{H}$, along with a collection $\mathcal{E}$ of ordered pairs of events $(E, D)$, such that:

- γ($h$) is in Ω, for all γ in Γ and $h$ in Ω; and

- for any event pair $(E,D)$ in $\mathcal{E}$ and any γ in Γ, if $E'$ and $D'$ are the inverse images of $E$ and $D$ under γ, then $(E',D')$ is also in $\mathcal{E}$, and $Pr_{E'}[D'] = Pr_E[D]$.

For example, in our random walk example (re-construed in the framework of spatially extended systems), let $\mathcal{E}$ be the set of all ordered pairs of events $(E, D)$ such that event $E$ exactly specifies the location of the particle at some time $t$, while event $D$ happens at some later time $t'$. Thus, $Pr_E[D]$ is the conditional probability that the token satisfies such-and-such property at time $t'$, *given* that it was at such-and-such location at time $t$. Let $\Gamma_{ST}$ be the monoid of all spatiotemporal translations of $S \times T$. If γ is any element of $\Gamma_{ST}$, then $\mathcal{E}$ is invariant under γ, and the conditional probability $Pr_E[D]$ is preserved by γ for any $(E,D)$ in $\mathcal{E}$, in the sense described above. Thus, the pair $(\Gamma_{ST}, \mathcal{E})$ is a partial symmetry monoid for the random-walk system. Crucially, the set $\mathcal{E}$ does not include pairs of the form $(\Omega, D)$, so we do not require unconditional probabilities of the form $Pr_\Omega[D]$ to be preserved by spatiotemporal translations.

Seen from this perspective, the transition probabilities of the random walk are "lawlike", because they are preserved by all the transformations in $\Gamma_{ST}$. In contrast, the initial probability distribution of the system is merely a "brute fact", since it is not preserved by any symmetries.

---

[73] The following argument does not depend on this assumption. Indeed, our argument would apply to *any* initial probability distribution for the token. Note that there is no such thing as a uniform probability distribution over the set of integers.

[74] To be precise: if $t$ is odd, and $t' = t–1$, then for any even $s$ between $–t'/2$ and $t'/2$, the probability that the token will be at spatial location $s$ at time $t$ is $2^{-t'}B((t'+s)/2, t')$, where $B$ is the binomial coefficient function. The formula for even times is similar, but more complicated.

For another example suppose that the temporally evolving (or spatially extended) system $(\Omega', \{Pr'_{E'}\}_{E' \subseteq \Omega'})$ is a *factor* of the system $(\Omega, \{Pr_E\}_{E \subseteq \Omega})$, via some factor map $\psi$, as described in Appendix A. For any event $E' \subseteq \Omega'$, let $\psi^{-1}(E')$ denote its inverse image under $\psi$ (here defined as a subset of $\Omega$). Then define $\mathcal{E} = \{(\psi^{-1}(E'), \psi^{-1}(D')): E', D' \subseteq \Omega'\}$. Let $\Gamma$ be a monoid of spatiotemporal symmetries of the factor system $(\Omega', \{Pr'_{E'}\}_{E' \subseteq \Omega'})$. The elements of $\Gamma$ might not be symmetries of the original system $(\Omega, \{Pr_E\}_{E \subseteq \Omega})$. However, they will be *partial* symmetries, with respect to the set $\mathcal{E}$. So $(\Gamma, \mathcal{E})$ is a partial symmetry monoid for $(\Omega, \{Pr_E\}_{E \subseteq \Omega})$. As already explained in Appendix A, one can greatly extend the scope of our framework by focusing attention on a factor system rather than the original system. We now see that this is a special case of the broader concept of a partial symmetry monoid.

However, partial symmetry monoids cannot accommodate another feature of many systems. To illustrate this, consider a temporally evolving system where time is a *finite* sequence of integers, e.g., $T = \{1,2,...,100\}$. For such a system, time translations are not even *well-defined*.[75] But in most such systems, we still want to say that the system obeys the same causal laws at all times, except perhaps at times 0 and 100. A similar problem arises in a spatially extended system where the space $S$ is bounded (e.g., a partial differential equation defined on a cube, with specified boundary conditions) or a finite set of points (e.g., a cellular automaton defined on a 100 × 100 grid, with specified boundary conditions). In such a system, spatial translations are not well-defined. But in most such systems, we still want to say that the system obeys the same causal laws everywhere in the "interior" of the spatial domain.

To solve this problem, we now introduce "local" symmetries. We begin with some preliminary definitions. Let $N$ be a subset of $S \times T$; heuristically, this represents some "region" of space-time. Extending our earlier terminology, we say that an event $E \subseteq \Omega$ *happens inside* $N$ if, for all histories $h$ and $h'$ in $\Omega$, if $h_N = h'_N$, then $h_N$ is in $E$ if and only if $h'_N$ is in $E$. Let $\Omega_N = \{h_N : h \text{ in } \Omega\}$; this is a collection of functions from $N$ into $X$. Let $N'$ be another subset of $S \times T$, and suppose $\gamma$ is a function from $N'$ into $N$. For any $h_N$ in $\Omega_N$, we define $\gamma(h_N)$ to be the function $h'_{N'}$ from $N'$ into $X$ defined by $h'_{N'}(n') = h_N[\gamma(n')]$, for all $n'$ in $N'$. Note that $h'_{N'}$ is not necessarily an element of $\Omega_{N'}$.

We now define a *local symmetry groupoid* of a spatially extended system to be a combination of three components:

- a collection $\mathcal{N}$ of subsets of $S \times T$ (called *neighbourhoods*);

- for each neighbourhood $N$ in $\mathcal{N}$, a set $\mathcal{E}_N$ of ordered pairs of events $(E, D)$ which happen inside $N$ (called *local events*); and

- for each pair of neighbourhoods $N$ and $N'$ in $\mathcal{N}$, a collection $\Gamma_{N,N'}$ of bijections from $N'$ into $N$ (called *local symmetries*).

---

[75] For example, suppose we try to define $\psi(t) = t+1$ for all $t$ in $T$; then $\psi(100)$ is not well-defined, because 101 is not an element of $T$.

We refer to the collection $\{\Gamma_{N,N'}: N, N' \in \mathcal{N}\}$ as a *groupoid* because it must satisfy two algebraic closure properties:

- For all neighbourhoods $M$ and $N$ in $\mathcal{N}$, and any local symmetries $\gamma$ in $\Gamma_{M,N}$, its inverse $\gamma^{-1}$ is in $\Gamma_{N,M}$.

- For all neighbourhoods $L$, $N$, and $M$ in $\mathcal{N}$, and all local symmetries $\alpha$ in $\Gamma_{L,M}$ and $\beta$ in $\Gamma_{M,N}$, the composition $\alpha \circ \beta$ is in $\Gamma_{L,N}$.

For any neighbourhoods $N$ and $N'$ in $\mathcal{N}$, and any $\gamma$ in $\Gamma_{N,N'}$, we call $\gamma$ a *local symmetry* because it must preserve the modal and probabilistic structure of the system in the following sense:

- For any $h$ in $\Omega_N$, its image $\gamma(h)$ is in $\Omega_{N'}$.

- For all event pairs $(E',D')$ in $\mathcal{E}_{N'}$, if $E$ and $D$ are the inverse images of $E'$ and $D'$ under $\gamma$, then $(E,D)$ is in $\mathcal{E}_N$, and $Pr_E[D] = Pr_E[D']$.

For example, suppose that $S = \{1,2,...,10\}$ and $T = \{1,2,...,100\}$. For any $s$ in $\{2,...,9\}$ and $t$ in $\{2,...,99\}$, let $N_{s,t}$ be the 3 × 2 "space-time rectangle" of the form $N_{s,t} = \{s-1, s, s+1\} \times \{t, t+1\}$. Let $\mathcal{N}$ be the set of all such space-time rectangles. For any $s$ and $s'$ in $\{2,...,9\}$, and any $t$ and $t'$ in $\{2,...,99\}$, if $N = N_{s,t}$ and $N' = N_{s',t'}$, then we define $\Gamma_{N,N'} = \{\gamma_{s',t' \to s,t}\}$, where $\gamma_{s',t' \to s,t}$ is the function from $N'$ into $N$ which sends each space-time point $(s_0, t_0)$ in $N'$ to the point $(s_0 - s' + s, t_0 - t' + t)$ in $N$. Then, with a suitable specification of the local event sets $\mathcal{E}_N$ for all $N$ in $\mathcal{N}$, we could construct a local symmetry groupoid for many of the spatially extended systems (such as cellular automata) that one might define on $S \times T$. However, a fully worked out example would be rather technically involved, and is beyond the scope of this paper; see Golubitsky, Pivato, and Stewart (2003) and Guay and Hepburn (2009).

Most of the ideas we have developed in this paper for the monoid of "full" symmetries can be generalized to partial symmetries and local symmetries. However, this is also beyond the scope of this paper.

**Appendix C: Spatial distance in quantum mechanical systems**

In Section 3.9, we proposed a definition of the *distance* between two regions $R$ and $R'$ in space, based on the minimum time duration required for a "signal" to travel from $R$ to $R'$. As we have already observed, this definition is not entirely satisfactory in systems where signals can travel at arbitrarily high speeds (such as classical mechanics). This is particularly problematic in quantum mechanics, for two reasons.

First, there is the well-known phenomenon of *entanglement*, where two particles, perhaps separated by a large spatial distance, can apparently correlate their behaviour. But in fact this is less of a problem than it first appears. Rather than interpreting entanglement as "spooky action at a distance", we can interpret it as a sign that we have not correctly specified the space $S$ for this spatially extended system. A three-dimensional quantum system with $n$ particles is not a collection of $n$ wave functions on a three-dimensional space; rather, it should be viewed as a *single* wave function on a $3n$-dimensional space. So we should define $S = \mathbf{R}^{3n}$. Even if two particles appear widely "separated" from our three-dimensional perspective, their joint location is described by a single "hump" of the wave function in a six-dimensional space.[76] From this perspective, the entangled behaviour of the two particles does *not* appear as a non-local phenomenon.

However, there is a more fundamental problem, which affects even a single-particle quantum system. Solutions to the Schrödinger equation on unbounded domains generally have *full support*: they give nonzero probability to every part of the space. This means, in effect, that the particle has a nonzero probability (albeit tiny) of "jumping" arbitrarily large distances through space.[77] Thus, no two regions of space are ever unreachable from one another in any time duration, no matter how short, and so the distance between any two regions will be zero, according to the definition given in Section 3.9.

To address this problem, we must introduce a slightly more nuanced version of "unreachability". Let $\varepsilon > 0$ be some small "error tolerance". Given three events $E$, $F$, and $G$ in $\Omega$, we say that $E$ and $G$ are ε-*conditionally independent given F*, if $1-\varepsilon < \Pr_F[E \cap G] / \Pr_F[E] \cdot \Pr_F[G] < 1+\varepsilon$. In other words, the conditional probability $\Pr_F[E \cap G]$ is "almost" the same as the product $\Pr_F[E] \cdot \Pr_F[G]$, which means that $E$ and $G$ are "almost" conditionally independent, given $F$. If $R$ and $R'$ are two regions of $S$, and $t > t_0$, then we say that $R'$ is ε-*unreachable* from $R$ in time $t$ if, for any $h$ in $\Omega$, any event which happens in $R'$ at time $t$ is ε-conditionally independent of any event which happens in $R$ at time $t_0$, given the event $[h_{R^c \times \{t_0\}}]$. If ε is small, this means that, *with very high probability*, a signal which originates in $R$ at time $t_0$ cannot reach $R'$ before time $t$. We then define the ε-*distance* between $R$ and $R'$ to be the supremum of the set of all $t$ such that $R'$ is ε-unreachable from $R$ in time $t$ (if this supremum exists).

By using a small but non-zero ε, we can thus define a non-trivial notion of ε-distance between different regions of space, even in a quantum-mechanical system. This measure of distance will obviously depend on the value of ε, but it will roughly approximate the "classical" notion of distance. However, a detailed development of this approach is beyond the scope of this paper.

---

[76] We are being slightly imprecise here; in quantum mechanics, particles do not even *have* precisely specified locations.
[77] Again, we are being somewhat imprecise in even ascribing a particular "location" to the particle. The Fourier transform of the wave function also has full support; this means that any velocity for the particle, no matter how large, has a tiny but non-zero probability of being realized.